\documentclass[aps,pre,reprint,twocolumn,showpacs,superscriptaddress,groupaddress,showkeys,floatfix,byrevtex]{revtex4-2}

\usepackage{graphicx}
\usepackage{dcolumn}
\usepackage{bm}

\usepackage{amsmath}
\usepackage{amsfonts,bm}
\usepackage{amssymb}
\usepackage[dvipsnames]{xcolor}
\usepackage{graphicx,color,subeqnarray}
\usepackage{dcolumn}
\usepackage{bm}
\usepackage{soul}

\begin{document}

\title{A Brownian cyclic engine operating in a viscoelastic active suspension}

\author{Carlos Antonio  \surname{Guevara-Valadez}}
\affiliation{Instituto de F\'isica, Universidad Nacional Aut\'onoma de M\'exico, Ciudad de M\'exico, C\'odigo Postal 04510, Mexico,}

\author{Rahul \surname{Marathe}}
\email{maratherahul@physics.iitd.ac.in}
\affiliation{Department of Physics, Indian Institute of Technology Delhi, New Delhi-110016, India.}

\author{Juan Ruben \surname{Gomez-Solano}}
\email{r\_gomez@fisica.unam.mx}
\affiliation{Instituto de F\'isica, Universidad Nacional Aut\'onoma de M\'exico, Ciudad de M\'exico, C\'odigo Postal 04510, Mexico,}

\date{\today}

\begin{abstract}
We investigate a model for a Stirling-like engine consisting of a passive Brownian particle confined by a harmonic potential and  interacting with a suspension of active Brownian particles that self-propel in a viscous solvent, which cyclically operates under isothermal conditions by means of temporal variations of the trap stiffness and the self-propulsion speed of the active particles. We derive an effective stochastic equation of motion of the trapped Brownian particle, which includes a friction memory kernel as well as thermal and active fluctuating forces due to its coupling with the active suspension, from which we analytically compute the efficiency of the engine in the quasi-static limit. We find that, on average, the engine is capable to produce mechanical work with an efficiency that depends on the interplay between the different time scales of the system, where the general effect of the ensuing viscoelasticity of the active suspension is to reduce the quasi-static efficiency of the
Brownian engine, as compared to the case of a system with instantaneous friction. Nevertheless, there are regions in the parameters space of the system where such memory effects are negligible in the performance of the engine, thus effectively behaving as in contact with an inert viscous bath working at two different temperatures related to the propulsion speed of the active particles.
\end{abstract}

\maketitle

\section{Introduction}

Recent advances in statistical mechanics have led to numerous applications of non-equilibrium systems in the mesoscopic realm, among which microscopic machines based on active matter, such as bacteria and synthetic self-propelled colloids, are currently attracting a great deal of attention \cite{Libchaber2000,Maggi14, Fodor16, steffenoni2016, Krishnamurty16, zakine2017, chaki2018, Arnab18, saha2019, Arnab20, Steffenoni20, Viktor20, Fodor20, Nroy21, Cates21EPL, albay2021, Slahiri21, Marathe22}. Unlike their passive counterpart, which require either cyclic temperature variations and baths at different temperatures \cite{Marathe07, Tschmidl08, Bechinger12,quinto2014, Rana14, Rana16, Gracia16, Robnagel16, Basu17, Viktor17, Seifert18, Vholubec18, Nascimento2021,Gomez21,JWRyu2021,miangolarra2022} or externally-driven asymmetric potentials \cite{rousselet94,lopez2008,haenggi2009} in order to produce a net power output or currents, active ratchets and active engines can exploit active matter \cite{ramaswamy2010,elgeti2015,bechinger2016,grosswasser2018}, either as a working substance or as an energy reservoir, 
to do work even under isothermal conditions \cite{Hiratsuka06, Campbell09, Leonardo10, Nori13, Zhang13, Ai14, Reichhardt17,Mishra19,Pietzonka19, Bisht20}. 
. 

In the last couple of decades, stochastic thermodynamics has been extremely useful in understanding the behavior of  passive and externally driven small-scaled systems in contact with a thermal bath, as it allows one to identify different thermodynamic quantities like heat, work and entropy production for single stochastic trajectories \cite{Sekimoto98, Seifret05, Seifert12, ciliberto2017}. Hence, it is quite natural to extend this theoretical framework to active systems, which are intrinsically subject to non-equilibrium fluctuations due the coupling between their specific mechanisms of energy conversion into directed motion and the physical properties of their environment. This includes deriving and validating different fluctuation theorems for active systems, thereby establishing second-law bounds on their entropy production~\cite{Debashish13, Debashish14, speck2016, Mandal17, pietzonka2018, Dabelow19, Szamel19, chaki2019, Goswami2021}, which have been essential in building stochastic models of active heat engines having efficiencies that respect Carnot-like constraints. For this it is required to specify conditions under which energy currents in the active system can be termed as thermodynamic heat. One way to achieve this is to map a non-equilibrium active system to an effective equilibrium situation, and identify the corresponding effective temperature \cite{Steffenoni20, Viktor20, Goswami2021, Szamel14, Petrelli2020, GomezSolano2020}, which in turn helps distinguish heat-to-work converters that respect the Carnot bound, from the work-to-work converters that do not follow the bounds set by the second law. It is important to note that such a mapping is not always possible, but for an overdamped system, like the one considered in the present manuscript, an effective equilibrium description can be provided. Such active engines can be realized by suspending passive (non-active) particles in a bath containing active particles (for example, run-and-tumble particles or active Brownian particles), or by suspending a few active particles in a bath of passive particles, and then confining the motion of the corresponding working substance by external potentials in order to extract mechanical work from the system. In such cases the statistical properties of single tracer particles or configurations of long individual macromolecules are altered due to the active entities present in the bath \cite{Libchaber2000,steffenoni2016,chaki2018, maes2015, ChakiJCM19, Goswami19, narinder2022}. This leads to diverse non-equilibrium phenomena in presence of confining potentials, such as enhanced diffusion and non-trivial noise correlations in the case of harmonic potentials~\cite{Maggi14}, and activated escape over energy barriers in multi-stable potentials \cite{Chaki2020,Ferrer2021}, whose first passage properties are of current interest~\cite{Woillez2019, Wexler20, militaru2021}. This type of active engines are potentially useful for the experimental realization of thermodynamic-like cycles with soft-matter systems by means of temporal variations of the confining potentials, e.g., with optical tweezers~\cite{Krishnamurty16,Nroy21,albay2021,Bechinger12,quinto2014}, whereas active baths with tunable properties have been recently implemented using thermally-sensitive bacteria~\cite{Krishnamurty16}, granular baths that are either magnetically driven \cite{tapia2020} or mechanically shaken \cite{Cheng2022}, etc. However, due to practical limitations and difficulties involved in controlling all active particles in the bath, an alternative is the artificial creation of active environments. Some of the methods recently used for this purpose are reservoir engineering by flashing optical traps~\cite{Nroy21}, random switching of the position of an optical trap~\cite{albay2021, Park20}, and external electrical noise~\cite{militaru2021}. Such setups may also help in generating non-trivial friction kernels and active noise correlations, which otherwise may be very difficult to tune naturally and remain limited to the single exponential type \cite{klapp2021}. 
In this context, analytical or numerical investigations of the working of cyclic active engines in terms of the efficiency under various physical conditions of the bath and the specific time-dependent protocol during the cycle, the effect of different time-scales like the cycle duration, the correlation times of the noise and of the friction kernel, the persistence time of the active particles, the fluctuations of the efficiency, and the distinct limiting cases of the engine operation, are of great present-day importance. Furthermore, finding the effects of viscoelasticity of the fluid environment~\cite{Gomez21}, non-Gaussianity~\cite{zakine2017}, and non-Markovianity of the active fluctuations~\cite{Marathe22, lee2020, Lee2021} on the performance of active engines  and improving the efficiency of active engines over that of passive ones \cite{Krishnamurty16, Arnab18, saha2019, Arnab20, Marathe22, lee2020, Lee2021, Nardini18, szamel2020, speck2022}, are also some important topics studied in the recent literature. The present article deals with some of these important aspects, as detailed below.

In almost all the theoretical studies on active heat engines mentioned above, a harmonically bound passive particle is described by an overdamped Langevin equation, where the trap strength is varied cyclically through a predefined time-dependent protocol to mimic the expansion and compression of the working substance. The active noise is considered as an extra additive colored noise together with the usual delta-correlated thermal noise. In such theoretical models, the non-trivial correlations of the active noise, which are usually chosen as mono-exponentially decay functions with a finite correlation time ~\cite{zakine2017,saha2019,Arnab20,Viktor20,Marathe22,lee2020,Lee2021}, break the fluctuation dissipation theorem and drive the system out of equilibrium. Though modelling of active noise correlations is phenomenological, it tends to describe very well the properties of the active noise and hence that of the suspended passive particle~\cite{Ye2020}. Motivated by above and some recent studies on collisional Brownian particle engines \cite{Fiore21} as well as classic works on the microscopic derivation of the generalized Langevin equation \cite{steffenoni2016,maes2015,ford1965, zwanzig1973}, in the present article we study a cyclic engine following a Stirling-like protocol, operating in a viscoelastic active suspension. To this end, unlike earlier studied models, we derive an active generalized Langevin equation from first principles, where memory friction and active noise with non-trivial correlations emerge naturally in the reduced description of a Brownian particle harmonically trapped in an active suspension. This system works as an engine even under isothermal conditions, where the self-propulsion speed of the active particles and the trap stiffness change periodically in time. The effects of different parameters like dissipation, the active noise correlation time, the ensuing viscoelasticity of the active suspension and the self-propulsion speed of its constituent particles, on the quasi-static efficiency of the engine, are studied in a completely analytical manner and compared with the passive counterpart.   

The paper is organized as follows. In Sec. \ref{sect:model} we describe the model and identify important parameters that influence the working of the active engine. In Sec. \ref{sect:variance} we explicitly calculate the variance of the position of the colloidal particle in a steady state. We require this to calculate the mean values of thermodynamic quantities like work, heat and the corresponding efficiency of the engine in the quasi-static limit of operation. In Sec. \ref{sect:activity} we introduce and study the Stirling-like protocol of our engine. We analytically calculate the work, heat and efficiency during different strokes of the engine. In this section we also study the efficiency of the engine in various parameter regimes. We summarize and conclude in Sec. \ref{sect:conclude}. In the Appendix \ref{app:derivGLE}, we provide a detailed microscopic derivation of the generalized Langevin equation considered in this paper.

\section{Model}\label{sect:model}

\begin{figure*}
    \centering
\includegraphics[width=1.95\columnwidth]{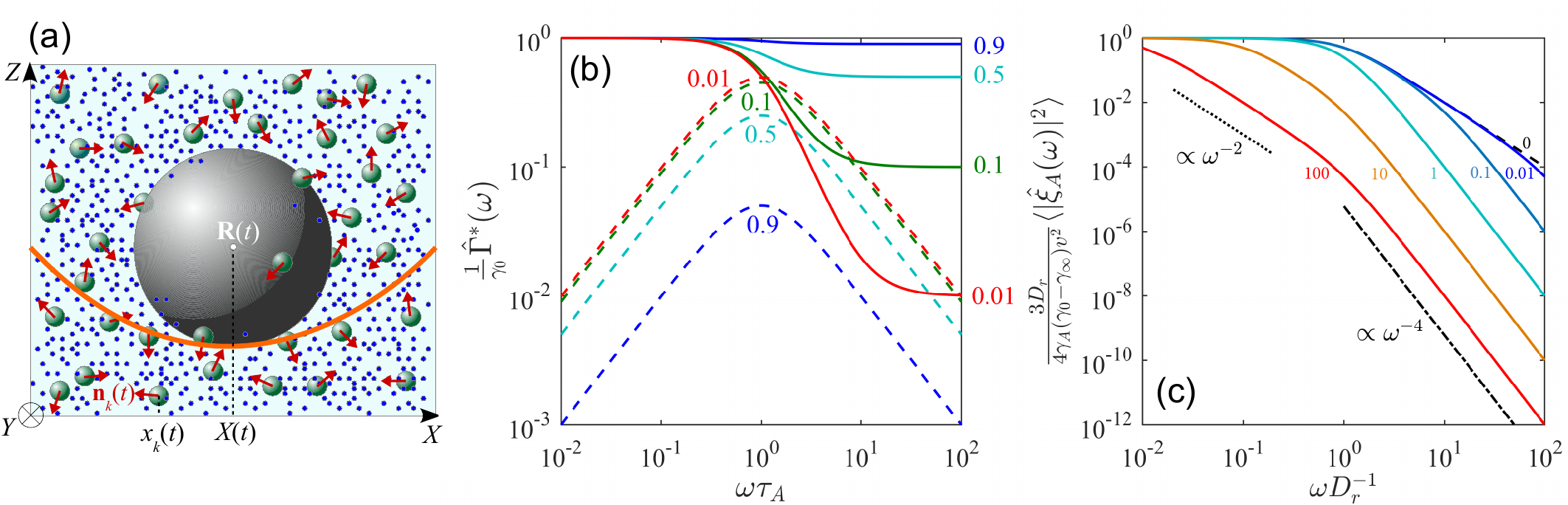}
\caption{(a) Sketch of a Brownian heat engine consisting of a passive Brownian particle (big gray sphere) confined by a harmonic potential (orange solid line) and embedded in an active suspension composed of passive solvent molecules (blue dots) at constant temperature $T$ and active Brownian particles (teal spheres) that can self-propel with a well defined orientation (red arrows) and a finite non-zero persistence time. The main coordinates defining the state of the system at time $t$ are also shown. (b) Real (solid lines) and imaginary part (dashed lines) of the complex frequency-dependent friction given by Eq. (\ref{eq:fourierkernel}) (normalized by $\gamma_0$) as a function of the frequency (normalized by $\tau_A^{-1}$) for different values of the parameter $\alpha$ defined in Eq. (\ref{eq:alpha}): $\alpha = 0.9, 0.5, 0.1, 0.01$. (c) Normalized power spectral density of the active noise given by Eq. (\ref{eq:powerspectrumactive}) (solid lines) as a function of the frequency (normalized by $D_r$) for different values of the parameter $\epsilon$ given by Eq. (\ref{eq:epsilon}). From right to left: $\epsilon = 0.01, 0.1, 1, 10, 100$. The dashed line represents the Lorentzian curve corresponding to the case $\epsilon = 0$. The dotted and dotted-dashed lines depict the dependencies $\propto \omega^{-2}$ and $\propto \omega^{-4}$, respectively.}\label{fig:1}
\end{figure*}

We consider a model for a stochastic heat engine consisting of a passive Brownian particle (PBP) trapped by a harmonic potential in a viscoelastic active suspension with tunable activity, which is capable of performing work at constant temperature by means of cyclic temporal variations of the trap stiffness and the activity of some of the surrounding particles. The PBP, whose position in three dimensions at time $t$ is described by the vector $\mathbf{R}(t) = [X(t),Y(t),Z(t)]$, is spherical (radius $a$) and surrounded by $N \gg 1$ identical spherical particles of radius $b \ll a$ that can become active by extracting energy from their surroundings. Both the Brownian particle and the $N$ active Brownian particles (ABPs) are embedded in a viscous solvent of viscosity $\eta_{\infty}$ that plays the role of a heat bath at constant temperature $T$, as schematized in Fig. \ref{fig:1}(a). The solvent is composed of $N_s \gg N$ particles whose size is much smaller than those of active particles. The PBP interacts with the ABPs via the potential $ V(\left\{ \mathbf{r}_i(t) \right\}, \mathbf{R}(t))$, where $\{{\mathbf{r}}_i(t)\}$ denotes the set of positions of the ABPs at time $t$, with $i = 1, \ldots, N$. A harmonic potential, $U(\mathbf{R}(t)) = \frac{1}{2} \kappa |\mathbf{R}(t)|^2$, acts directly on the PBP, thus confining its stochastic motion. Therefore, in the overdamped limit, the position of the PBP evolves in time according to the Langevin equation
\begin{equation}\label{eq:LangevinBrownianengine}
 \gamma_{\infty}\frac{d}{dt}\mathbf{R}(t) = -\kappa \mathbf{R}(t) - \nabla_{\mathbf{R}} V(\left\{ \mathbf{r}_i(t) \right\},\mathbf{R}(t)) + \bm{\zeta}_{\infty}(t).
\end{equation}
where $\gamma_{\infty} = 6\pi \eta_{\infty} a$ is the friction coefficient experienced by the PBP in the viscous solvent, whereas $\bm{\zeta}_{\infty}(t)$ is a Gaussian white noise that mimics the effect of the thermal collision of the $N_s$ solvent molecules. The mean and autocorrelation function of $\bm{\zeta}_{\infty} (t)$ are
\begin{eqnarray}\label{eq:white_noise1}
    \langle \bm{\zeta}_{\infty}(t) \rangle & = & \mathbf{0},\nonumber\\
    \langle \bm{\zeta}_{\infty}(t) \otimes \bm{\zeta}_{\infty}(s) \rangle & = & 2k_B T \gamma_{\infty}\delta(t-s) \mathbb{I},
\end{eqnarray}
respectively, where $\otimes$ stands for the dyadic product, $\delta(t-s)$ is the Dirac delta function, and $\mathbb{I}$ represents the identity tensor. On the other hand, the three-dimensional motion of the ABPs at time $t$ is described by the stochastic model \cite{bechinger2016}
\begin{equation}\label{eq:ABM_model}
    \frac{d}{dt} \mathbf{r}_{i}(t) = v \mathbf{n}_i(t) - \mu \nabla_{\mathbf{r}_i} V(\left\{ \mathbf{r}_i(t) \right\}, \mathbf{R}(t)) +  \bm{\chi}_i(t),
\end{equation}
where $v$ is the self-propulsion speed of the active particles, $\mathbf{n}_i(t)$ is a unit vector that represents the instantaneous orientation of the $i-$th ABP particle, $\mu = \gamma_A^{-1}$ is the  mobility of each ABP, which is given by the inverse of its friction coefficient in the solvent, $\gamma_A = 6\pi \eta_{\infty} b$, and $\bm{\chi}_i(t)$ is a Gaussian white noise with mean and autocorrelation function 
\begin{eqnarray}\label{eq:whitenoise2}
    \langle \bm{\chi}_i(t) \rangle & = & \mathbf{0},\nonumber\\
    \langle \bm{\chi}_i(t) \otimes \bm{\chi}_{j}(s) \rangle & = & 2k_B T \mu \delta_{ij}\delta(t-s) \mathbb{I},
\end{eqnarray}
respectively, with $\delta_{ij}$ the Kronecker delta. The particle orientations have the following correlation functions
\begin{equation}\label{eq:persist_orientation}
    \langle \mathbf{n}_i(t) \cdot \mathbf{n}_j(s) \rangle = \delta_{ij} \exp \left( -D_r |t-s|\right)
\end{equation}
which capture the effect of rotational diffusion on the persistence of the instantaneous self-propulsion velocity, $v {\mathbf{n}}_i(t)$. In Eq. (\ref{eq:persist_orientation}), $D_r$ represents the rotational diffusion coefficient, whose inverse $D_r^{-1}$ sets the typical persistence time during which each active particle exhibits a rather directed motion with a characteristic length-scale $v D_r^{-1}$.
For the sake of simplicity, in Eqs. (\ref{eq:ABM_model}) and (\ref{eq:persist_orientation}), we assume that the ABPs do not interact with each other but only with the bigger PBP, which is sufficient to capture a non-trivial viscoelastic coupling with the surroundings.
Moreover, we consider that $ V(\{{\mathbf{r}}_i\}, {\mathbf{R}})$ can be modeled as a sum of harmonic interactions between the Brownian engine particle and the active ones with elastic constant $k$, which is an approximation commonly used in the derivation of generalized Langevin equations \cite{ford1965,zwanzig1973}. Therefore, as shown in Appendix \ref{app:derivGLE}, by following the approach proposed by~\cite{steffenoni2016,maes2015} based on nonequilibrium linear-response theory in the limit of a weak interaction, the stochastic time evolution of a single Cartesian coordinate of the PBP position, \emph{e.g.}, $X(t)$, can be effectively described by the generalized Langevin equation
\begin{equation}\label{eq:GLEactive}
    \int_{-\infty}^t ds \, \Gamma(t-s) \frac{d}{ds} X(s) = - \kappa X(t)  + \xi(t).
\end{equation}
In the derivation of Eq. (\ref{eq:GLEactive}), the strength of the interaction potential has been chosen as $k \ll N^{-1} \kappa$, in such a way that after coarse-graining the ABP dynamics, the resulting stiffness of the total confining potential acting on the PBP, $(1-Nk/\kappa)\kappa\approx \kappa$, remains  independent of $k$, see Appendix~\ref{app:derivGLE}. In addition, the degrees of freedom of the ABPs have been averaged out, where the interaction force with the active suspension results in a friction memory kernel $\Gamma(t-s)$ that is given by
\begin{equation}\label{eq:totalkernel}
	\Gamma(t-s) =   2\gamma_{\infty} \delta(t-s)  + \frac{\gamma_0 - \gamma_{\infty}}{\tau_A} \exp \left( - \frac{t-s}{\tau_A} \right),
\end{equation} 
and a stochastic force $\xi(t) \equiv \xi_T(t) + \xi_A(t)$, where $\xi_T(t)$ and $\xi_A(t)$ are Gaussian noises with zero mean, \emph{i.e.}, $\langle \xi_T(t) \rangle = \langle \xi_A(t) \rangle = 0$, and correlations
\begin{widetext}
\begin{eqnarray}
	\langle  \xi_T(t)  \xi_T(s) \rangle &  = & k_B T \Gamma(|t-s|), \label{eq:thermalnoise} \\
	\langle \xi_A(t) \xi_A(s) \rangle & = & \frac{\gamma_A(\gamma_0 - \gamma_{\infty})v^2}{3 \left( 1 - D_r^2 \tau_A^2 \right)}\left[  \exp \left( -D_r |t-s| \right) -D_r \tau_A \exp\left( -\frac{|t-s|}{\tau_A} \right) \right], \label{eq:activenoise}\\
	\langle \xi_T(t) \xi_A(s) \rangle & = & 0. \label{eq:thermalactive}
\end{eqnarray}
\end{widetext}
On the right-hand-side of Eq. (\ref{eq:totalkernel}), the first term is directly related to the instantaneous friction exerted by the solvent on the PBP, whereas the second term results from the retarded viscoelastic response of the active suspension due to the interaction between the PBP and the ABPs. In Eqs. (\ref{eq:GLEactive}) and  (\ref{eq:thermalnoise})-(\ref{eq:activenoise}), we have split the stochastic force exerted by the active suspension on the PBP into a contribution originating from the thermal motion of the ABPs and the solvent particles, $\xi_T(t)$, and a contribution entirely due to the non-equilibrium self-propulsion of the ABPs, $\xi_A(t)$. Moreover, we have expressed the parameters $k$ and $\mu = \gamma_A^{-1}$ that are related to the dynamics of the ABPs in terms of effective parameters that describe the viscoelastic coupling of the PBP with the active suspension.
To this end, we define $\tau_A = \frac{\gamma_A}{k}$, which represents the characteristic time-scale of an ABP when interacting with the PBP and can thus be interpreted as the structural relaxation time of the active suspension. Furthermore, we have considered that,
in the zero-deformation-rate limit, the friction coefficient of the PBP
can be written as
\begin{equation}\label{eq:zerofriction}
 \gamma_0  =  \int_0^{\infty} dt' \, \Gamma(t')
  =   \gamma_{\infty} + \frac{N}{\mu},
 \end{equation}
where we have used the explicit functional form of the memory kernel derived in Appendix \ref{app:derivGLE}, see Eqs. (\ref{eq:activekernel}) and (\ref{eq:totalfrictionkernel}). Eq. (\ref{eq:zerofriction}) implies that $\frac{N}{\mu} = N\gamma_A = \gamma_0 - \gamma_{\infty}$, and is consistent with the fact that the total energy dissipation of the trapped PBP happens both directly through the solvent with friction coefficient $\gamma_{\infty}$, and indirectly through the interaction with the $N$ active particles, each one with friction coefficient $\frac{1} {\mu} = \gamma_A$.
Note that the memory kernel in Eq.~(\ref{eq:totalkernel}) with the effective coupling parameters $\gamma_0$, $\gamma_{\infty}$ and $\tau_A$ is similar to that for the friction experienced by a bead in a Stokes-Oldroyd B fluid, which represents the minimal model of linear viscoelasticity of complex fluids incorporating both the transient elastic response of the fluid microstructure and the dissipation of the solvent~\cite{paul2018}. Furthermore, according to Eqs.~(\ref{eq:thermalnoise}) and (\ref{eq:activenoise}), the noise $\xi(t) = \xi_T(t) + \xi_A(t)$ in the generalized Langevin equation (\ref{eq:GLEactive}) does not satisfy the fluctuation-dissipation theorem of the second kind if $v > 0$ since $\langle \xi(t) \xi(s) \rangle \neq  k_B T \Gamma\left( | t-s| \right)$, thus reflecting the breakdown of detailed balance in the system due to the non-equilibrium activity of suspension. Thermal equilibrium is correctly recovered only for $v = 0$, in which case $\langle \xi(t) \xi(s) \rangle  = \langle \xi_T(t) \xi_T(s) \rangle = k_B T \Gamma\left( | t-s| \right)$.

For $D_r \tau_A \ll 1$, \emph{i.e.}, for $b \ll \frac{k} {6\pi \eta_{\infty} D_r}$ (vanishingly small ABPs), the generalized Langevin equation (\ref{eq:GLEactive}) reduces to 
\begin{equation}\label{eq:modifiedLangevin}
	\gamma_0 \frac{d}{dt}X(t) = -\kappa X(t) + \xi_T(t) + \xi_A(t),
\end{equation}
where $\gamma_0$ is the viscous drag coefficient of the PBP that gives rise to an instantaneous friction with the bath,  $\xi_T(t)$ is a Gaussian white noise of zero mean and autocorrelation $\langle \xi_T(t) \xi_T(s) \rangle = 2k_B T \gamma_0 \delta(t-s)$, whereas $\xi_A(t)$ is a noise of zero mean, and mono-exponentially decaying autocorrelation function
\begin{equation}\label{eq:autocorrexp}
    \langle \xi_A(t) \xi_A(s) \rangle = \frac{1}{3}\gamma_A (\gamma_0 - \gamma_{\infty}) v^2 e^{-D_r|t-s|}.
\end{equation}
We point out that phenomenological models based on Langevin equations similar to Eq. (\ref{eq:modifiedLangevin}), with exponentially correlated noises in the form of Eq.~(\ref{eq:autocorrexp}), have been extensively used to investigate the operation of active engines~\cite{saha2019,Arnab20,Viktor20,Marathe22}. Furthermore, stochastic engines  driven by active Ornstein-Uhlenbeck noises, whose autocorrelations are exponential decay functions similar to Eq.~(\ref{eq:autocorrexp}) with a multiplicative term proportional to the inverse of the rotational diffusion time, $1/D_r^{-1} = D_r$, have also been studied in previous theoretical works in the case of memoryless friction~\cite{zakine2017,Marathe22,lee2020,Lee2021}. Moreover, a particular version of Eq. (\ref{eq:modifiedLangevin}) driven by non-thermal white noise [$D_r \rightarrow \infty$ in Eq. (\ref{eq:autocorrexp})] was recently applied to investigate the performance of a colloidal Stirling engine in contact with a water reservoir at a single constant temperature, where a confining potential and a virtual (active) temperature varying in time according to a Stirling cycle was experimentally implemented by means of optical tweezers \cite{albay2021}. Therefore, the non-Markovian stochastic model given by Eqs. (\ref{eq:GLEactive})-(\ref{eq:thermalactive}), which is derived in the present article from first principles, provides the simplest non-trivial description of the motion of a passive Brownian particle suspended in both  viscous and viscoelastic active suspensions and is consistent with the aforementioned phenomenological models of mesoscopic heat engines subject to active fluctuations.

It should be noted that ABP mobility, $\mu = \gamma_A^{-1}$ affects both the viscosity difference $\gamma_0 - \gamma_{\infty} = N\gamma_A$ and the relaxation time $\tau_A = \frac{\gamma_A}{k}$, which are viscoelastic parameters directly involved in the expressions for the friction memory kernel, $\Gamma(t)$, and the active-noise autocorrelation, $\langle \xi_A(t) \xi_A(0) \rangle$, see Eqs. (\ref{eq:totalkernel}) and (\ref{eq:activenoise}). 
To illustrate its main effects on the behaviors of $\Gamma(t)$ and $\langle \xi_A(t) \xi_A(0) \rangle$, we compute their Fourier transforms, defined as $\hat{f}(\omega) = \int_{-\infty}^{\infty} dt\, e^{-\mathrm{i}\omega t} f(t)$. In the case of the friction memory kernel, the complex conjugate of the Fourier transform of Eq.~(\ref{eq:totalfrictionkernel}), which can be interpreted as a complex frequency-dependent friction coefficient, reads
\begin{equation}\label{eq:fourierkernel}
    \hat{\Gamma}^{*}(\omega) = \gamma_0 \left\{ \left( \alpha + \frac{1-\alpha}{1+\omega^2\tau_A^2}\right) + \mathrm{i} \left[ \frac{(1-\alpha)\omega \tau_A}{1+\omega^2 \tau_A^2} \right] \right\}.
\end{equation}
In Eq. (\ref{eq:fourierkernel}), we have introduced the dimensionless parameter $\alpha$, defined as
\begin{equation}\label{eq:alpha}
	\alpha = \frac{\gamma_{\infty}}{\gamma_0},
\end{equation}
whose possible values are contained in the interval $0 < \alpha < 1$, and represents the ratio of the high- and low-frequency friction coefficients experienced by the PBP, $\hat{\Gamma}^{*}(\omega\rightarrow \infty) = \gamma_{\infty}$ and $\hat{\Gamma}^{*}(\omega\rightarrow 0) = \gamma_{0}$, respectively. For a dilute suspension of ABPs or one with vanishingly small values of $\gamma_A$ such that $N\gamma_A \ll \gamma_{\infty}$, we have $\alpha \rightarrow 1$, whereas for a dense active suspension or one with sufficiently large values of $\gamma_A$ such that $N\gamma_A \gg \gamma_{\infty}$, $\alpha \rightarrow 0$.
In Fig.~\ref{fig:1}(b) we plot the real and the imaginary parts of $\hat{\Gamma}^*(\omega)$, $\mathfrak{Re}\left[\hat{\Gamma}^*(\omega)\right]$ and $ \mathfrak{Im}\left[\hat{\Gamma}^*(\omega)\right]$, respectively, as a function of the normalized frequency $\omega \tau_A$ for different values of $\alpha$. In all cases, 
$\mathfrak{Im}\left[\hat{\Gamma}^*(\omega) \right]$, which is related to the energy storage of the system, exhibits a peak around $\omega = \tau_A^{-1}$, whose maximum is $\mathfrak{Im}\left[\hat{\Gamma}^*(\tau_A^{-1}) \right] = \frac{1}{2}(1-\alpha)\gamma_0$, whereas $\mathfrak{Re}\left[\hat{\Gamma}^*(\omega) \right]$, which quantifies the energy dissipation, decreases  monotonically from $\gamma_0$ to $\gamma_{\infty}$ with increasing values of $\omega$. We observe that for $\alpha$ close to 1, $\mathfrak{Im}\left[\hat{\Gamma}^*(\omega) \right]$ is orders of magnitude smaller than $\mathfrak{Re}\left[\hat{\Gamma}^*(\omega) \right]$ for all frequencies, the latter depending very weakly on $\omega$, as shown in Fig.~\ref{fig:1}(b) for $\alpha = 0.9$. As $\alpha$ decreases, $\mathfrak{Re}\left[\hat{\Gamma}^*(\omega) \right]$ has a more pronounced decay as a function of $\omega$, whereas the peak of $\mathfrak{Im}\left[\hat{\Gamma}^*(\omega) \right]$ around $\omega = \tau_A^{-1}$ increases and gets closer to the values of $\mathfrak{Re}\left[\hat{\Gamma}^*(\omega) \right]$, as shown in Fig.~\ref{fig:1}(b) for $\alpha = 0.5$. On the other hand, if $\alpha$ is sufficiently small, specifically if $\alpha < 3-2\sqrt{2} \approx 0.1716$, a frequency interval where the $\mathfrak{Im}\left[\hat{\Gamma}^*(\omega) \right] > \mathfrak{Re}\left[\hat{\Gamma}^*(\omega) \right]$ develops, as observed for $\alpha = 0.1$. A further decrease in $\alpha$ leads to a broadening of such a frequency interval, as verified in Fig. \ref{fig:1}(b) for $\alpha = 0.01$. The emergence of such an interval in Fourier-frequency domain indicates the transient regime in which elastic effects are dominant with respect to viscous dissipation in the drag force on the PBP in the active suspension, and cannot be neglected in the performance of the Brownian engine. Moreover, in the case of the autocorrelation function of the active noise force $\xi_A(t)$, the Fourier transform of Eq.~(\ref{eq:activenoise}) leads to the corresponding one-sided power spectral density, which is defined for $\omega \ge 0$, and has the following expression
\begin{equation}\label{eq:powerspectrumactive}
    \langle |\hat{\xi}_A(\omega)|^2 \rangle =  \frac{4\gamma_A(\gamma_0-\gamma_{\infty}) v^2}{3D_r(1+\omega^2D_r^{-2})(1+\epsilon^2 \omega^2 D_r^{-2})},
\end{equation}
where we have defined the ratio of the relaxation time of the active suspension, $\tau_A$, and the rotational diffusion time of the ABPs, $D_r^{-1}$, by means of the dimensionless parameter $\epsilon$
\begin{equation}\label{eq:epsilon}
	\epsilon = \frac{\tau_A}{D_r^{-1}} = D_r \tau_A \ge 0.
\end{equation}
In Fig.~\ref{fig:1}(c), we plot the power spectral density of the active noise given by Eq.~(\ref{eq:powerspectrumactive}) (normalized by $\frac{3D_r}{4\gamma_A(\gamma_0 - \gamma_{\infty})v^2}$) as a function of the Fourier frequency $\omega$ (normalized by $D_r$) for different values of $\epsilon$. For comparison, we also plot as a dashed line the power spectrum in the case of ABPs with vanishingly small $\gamma_A$, which correspond to the well-studied case with $\tau_A \rightarrow 0$, \emph{i.e.}, $\epsilon \rightarrow 0$, where the spectrum is Lorentzian according to Eq.~(\ref{eq:powerspectrumactive}). Interestingly, we find  the emergence of a transition to a high-frequency regime with increasing values of $\epsilon$, where $\langle |\hat{\xi}_A(\omega)|^2 \rangle \propto \omega^{-4}$, which is qualitatively distinct from the high-frequency Lorentzian behavior, $\langle |\hat{\xi}_A(\omega)|^2 \rangle \propto \omega^{-2}$ , for $\epsilon \rightarrow 0$. Therefore, it is expected that the performance of the Brownian engine will also be strongly affected by $\epsilon$, since high-frequency components of the active noise are more and more suppressed with increasing values of this parameter.

\begin{figure}
\includegraphics[width=1\columnwidth]{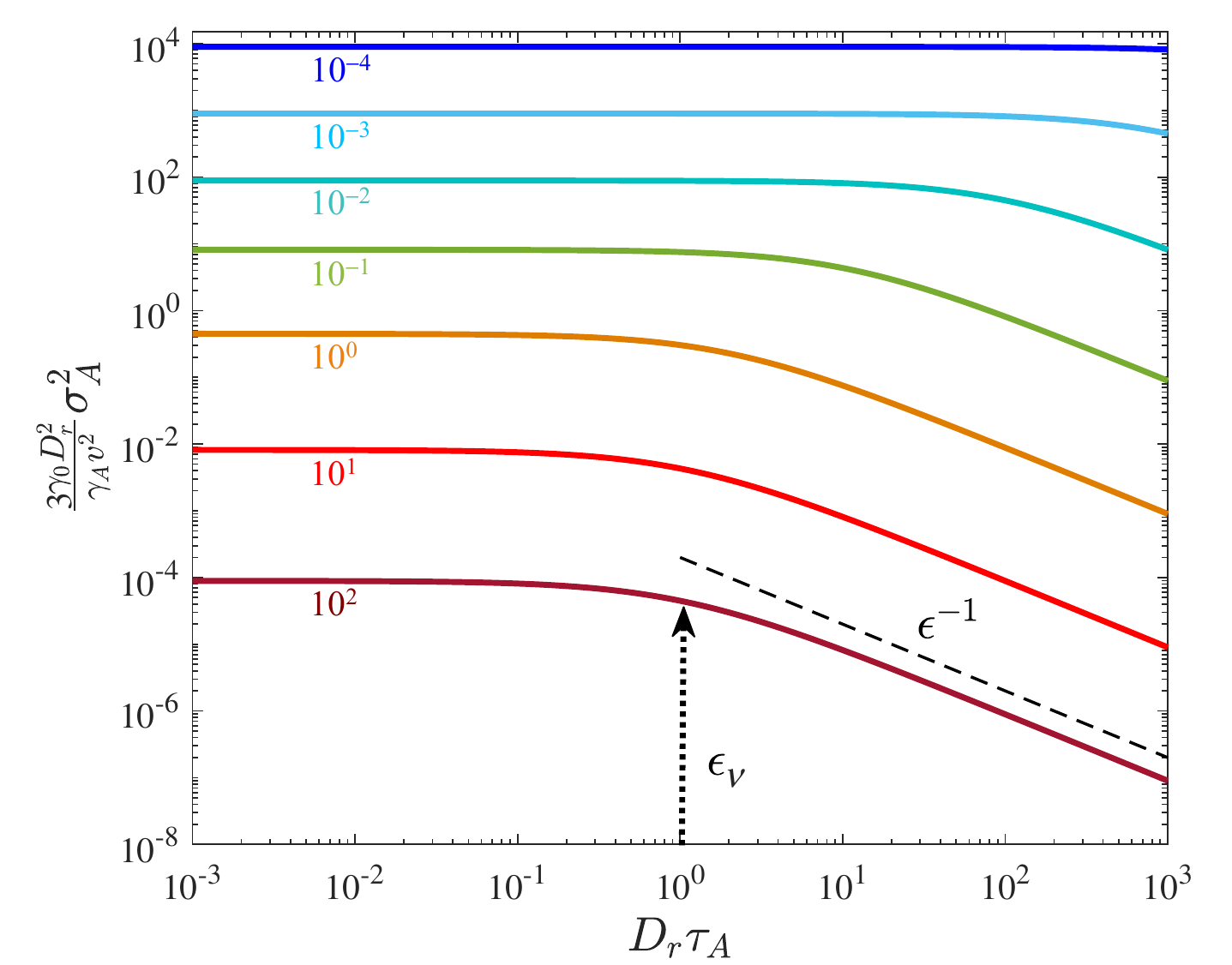}
\caption{Active component of the variance of the Brownian particle position (normalized by the squared length-scale $\frac{\gamma_A v^2}{3\gamma_0 D_r^2}$) as a function of the ratio $\epsilon$ between the viscoelastic relaxation time $\tau_A$ and the rotational diffusion time $D_r^{-1}$ of the ABPs with $\alpha = 0.1$ and for different values of the ratio $\nu$ between the rotational diffusion time and the largest viscous relaxation time $\gamma_0/\kappa$. From top to bottom: $\nu = 10^{-4}, 10^{-3}, 10^{-2}, 10^{-1}, 10^0, 10^1, 10^2$. The dotted arrow depicts the value $\epsilon_{\nu}$ for $\nu = 10^2$ above which the monotonic decrease in the variance, $\sigma_A^2 \propto \epsilon^{-1}$, becomes apparent (dashed line).}\label{fig:2}
\end{figure}

\section{Steady-state positional variance of the trapped PBP in the active suspension}\label{sect:variance}

From Eq. (\ref{eq:GLEactive}), the variance of the particle position, $\sigma^2$, in a non-equilibrium steady-state at constant $\kappa$, $T$ and $v$, reads
\begin{widetext}
\begin{equation}\label{eq:varianceX}
	\sigma^2 = \lim_{\theta \rightarrow \infty} \int_0^{\theta}  dt' \int_0^{\theta} dt'' \, I(\theta - t') I(\theta - t'') \left[ \langle \xi_T(t') \xi_T(t'') \rangle + \langle \xi_T(t') \xi_A(t'') \rangle + \langle \xi_A(t') \xi_T(t'') \rangle + \langle \xi_A(t') \xi_A(t'')\rangle  \right],
\end{equation}
\end{widetext}
where $I(t)$ is the inverse Laplace transform of the function
\begin{equation}\label{eq:Laplacekernel}
	\widetilde{I}(s) = \frac{1}{\kappa + s \widetilde{\Gamma}(s)},
\end{equation}
with 
\begin{equation}\label{eq:laplace_kernel}
    \widetilde{\Gamma}(s) = \int_0^{\infty}dt \, e^{-st} \Gamma(t) = \gamma_{\infty} + \frac{\gamma_0 - \gamma_{\infty}}{1 + \tau_A s}
\end{equation}
the Laplace transform of the friction memory kernel (\ref{eq:totalkernel}). Using Eqs. (\ref{eq:thermalnoise}), (\ref{eq:activenoise}), (\ref{eq:thermalactive}), (\ref{eq:varianceX}), and (\ref{eq:Laplacekernel}), an explicit expression for $\sigma^2$ can be derived in a straightforward manner
\begin{widetext}
\begin{eqnarray}
	\sigma^2 &  = & \sigma_T^2 + \sigma_A^2,\nonumber \\
		& = & \frac{k_B T}{\kappa}  + \frac{\left( 1 + \frac{\gamma_{\infty}}{\gamma_0}D_r \tau_A + \frac{\tau_A\kappa}{\gamma_0} \right)\gamma_A \left(\gamma_0-\gamma_{\infty} \right) v^2}{3 \gamma_0 D_r \kappa \left( 1+\frac{\tau_A \kappa}{\gamma_0} \right) \left[ 1 + \frac{\gamma_{\infty}}{\gamma_0}D_r \tau_A + \left(1+ D_r \tau_A \right) \frac{\kappa}{\gamma_0 D_r} \right]}.\label{eq:varianceposition}
\end{eqnarray}
\end{widetext}
In Eq. (\ref{eq:varianceposition}), $\sigma_T^2 \equiv \frac{k_B T}{\kappa}$ corresponds to the thermal contribution to the variance of $X(t)$, whereas $\sigma_A^2 = \sigma^2  - \sigma_T^2$ originates from the activity of the ABPs suspension. As can be seen in (\ref{eq:varianceposition}), unlike the thermal contribution, $\sigma_A^2$ depends on the interplay between the different relaxation processes of the system, specifically on the ratio between the time-scales $D^{-1}$, $\tau_A$ and $\gamma_0 / \kappa$, the latter being the largest relaxation time of the particle by viscous dissipation in the harmonic potential. If $D_r >0$, the active contribution to the variance can be expressed as
\begin{equation}\label{eq:varact}
	\sigma_A^2 = \frac{\gamma_A v^2}{3\gamma_0 D_r^2}\frac{\left( 1 - \alpha \right) \left[ 1 + (\alpha + \nu) \epsilon \right]}{\nu \left( 1 + \nu \epsilon \right) \left[ 1+\nu + \left( \alpha + \nu \right)\epsilon \right]},
\end{equation}
where $\alpha$ and $\epsilon$ are defined in Eqs.~(\ref{eq:alpha}) and (\ref{eq:epsilon}), respectively, and $\nu$ denotes the ratio of rotational diffusion time and the viscous dissipation time $\gamma_0/\kappa$
\begin{equation}\label{eq:nu}
	\nu = \frac{D_r^{-1}}{\frac{\gamma_0}{\kappa}} = \frac{\kappa}{\gamma_0 D_r} > 0.
\end{equation}
In Fig.~\ref{fig:2}, we plot the dependence described by Eq.~(\ref{eq:varact}) of $\sigma_A^2$  (normalized by the squared length-scale $\frac{\gamma_A v^2}{3\gamma_0 D_r^2}$) on $\epsilon$ for different values of $\nu$ ranging from $10^{-4}$ to $10^2$. We observe that, for sufficiently small $\epsilon$,  the viscoelastic relaxation time $\tau_A$ does not have a significant effect on the active component of the variance, whose value remains rather constant and very close to that for a Brownian particle in contact with an active bath with instantaneous friction kernel, $\Gamma(t-s) = 2 \gamma_0 \delta(t-s)$, namely $\sigma_A^2 = \frac{\gamma_A v^2}{3\gamma_0 D_r^2}\frac{1-\alpha}{\nu(1+\nu)}$ for $\tau_A =  0$, see \cite{zakine2017}. However, a monotonic decrease of the active component of the variance due to viscoelastic memory effects of the active suspension becomes apparent at sufficiently large $\epsilon$, where the behavior $\sigma_A^2 \propto \epsilon^{-1}$ is observed for $\epsilon$ larger than a certain value $\epsilon_{\nu}$ that depends on $\nu$. For a very weak trapping potential (sufficiently small $\nu$), $\epsilon_{\nu} \approx \nu^{-1}$, whereas  for a strongly confining potential (large $\nu$), $\epsilon_{\nu} \approx \frac{1+\nu}{\alpha + \nu} \rightarrow_{\nu \rightarrow \infty}  1$. Thus, viscoelastic memory effects hinder the active fluctuations undergone by a strongly trapped PBP (with $\kappa \gg \gamma_0 D_r$) with respect to the memoryless case ($\tau_A = 0$) whenever the rotational diffusion time of the ABPs in the suspension is typically shorter than the memory friction time $\tau_A$, \emph{i.e.}, if $D_r^{-1}  \lesssim \tau_A$.

\section{Stirling-like cycle under temporal variation of the suspension activity}\label{sect:activity}

\begin{figure}
    \centering
\includegraphics[width=0.95\columnwidth]{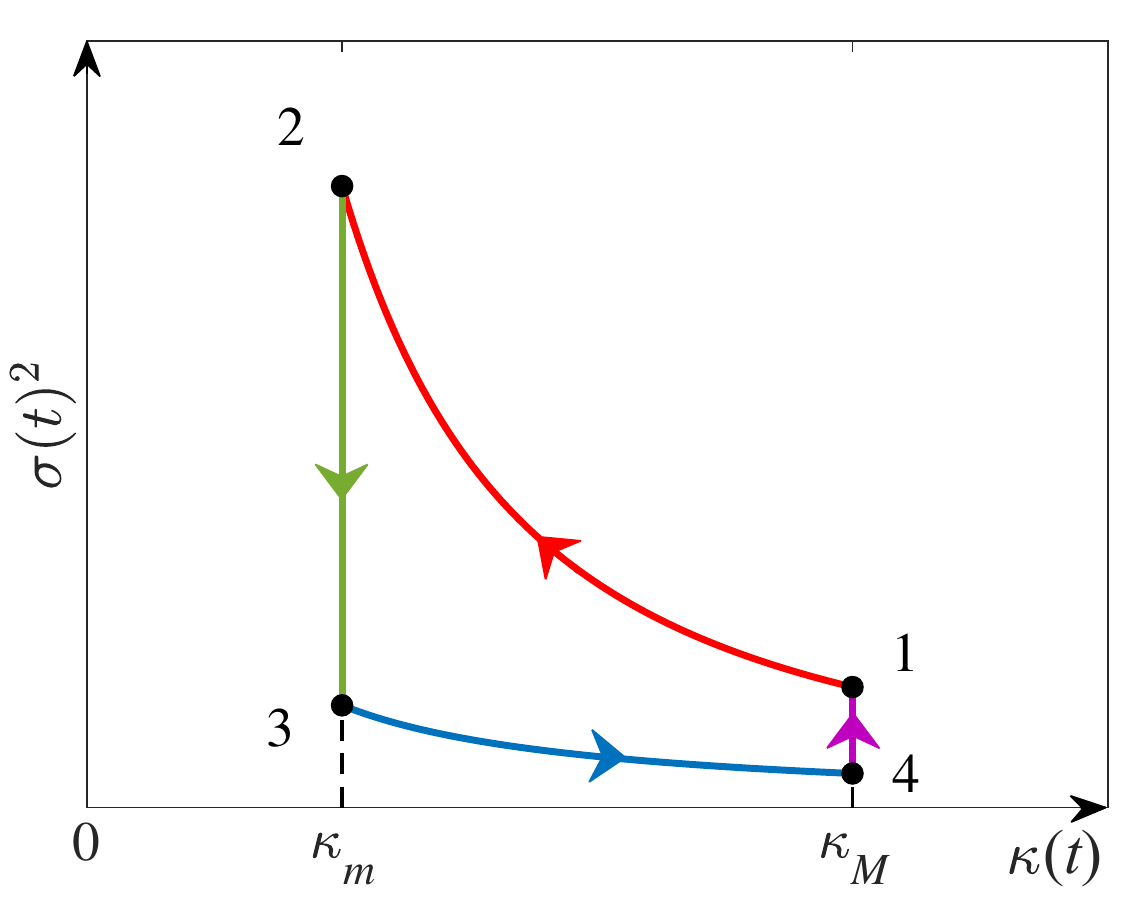}
\caption{Schematic representation in a positional variance-stiffness diagram of a Stirling-like cycle of duration $\tau$ performed by the Brownian heat engine depicted in Figure \ref{fig:1}(a) by means of the temporal variation of the stiffness $\kappa(t)$ of the trapping harmonic potential and of the propulsion speed of the active particles in the bath, $v(t)$. At time $0 \le t < \tau/2$, the trap stiffness is linearly decreased from $\kappa_M$ to $\kappa_m < \kappa_M$ while keeping the propulsion speed at $v(t) = v_0 > 0$ (step $1\rightarrow 2$). At $t = \tau/2$, the propulsion speed is suddenly switched off, $v(t) = 0$, with $\kappa(t =\tau/2) = \kappa_m$ (step $2\rightarrow 3$). Afterwards, the propulsion speed is kept at $v(t) = 0$ for $\tau/2 < t < \tau$, while linearly increasing the trap stiffness from $\kappa_m$ to $\kappa_M$ (step $3\rightarrow 4$). The cycle is completed at $t = \tau$, at which the propulsion speed is switched on again, $v(t) = v_0$ with $\kappa(t =\tau) = \kappa_M$ (step $4 \rightarrow 1$).}\label{fig:3}
\end{figure}

\begin{figure*}
    \centering
\includegraphics[width=1.95\columnwidth]{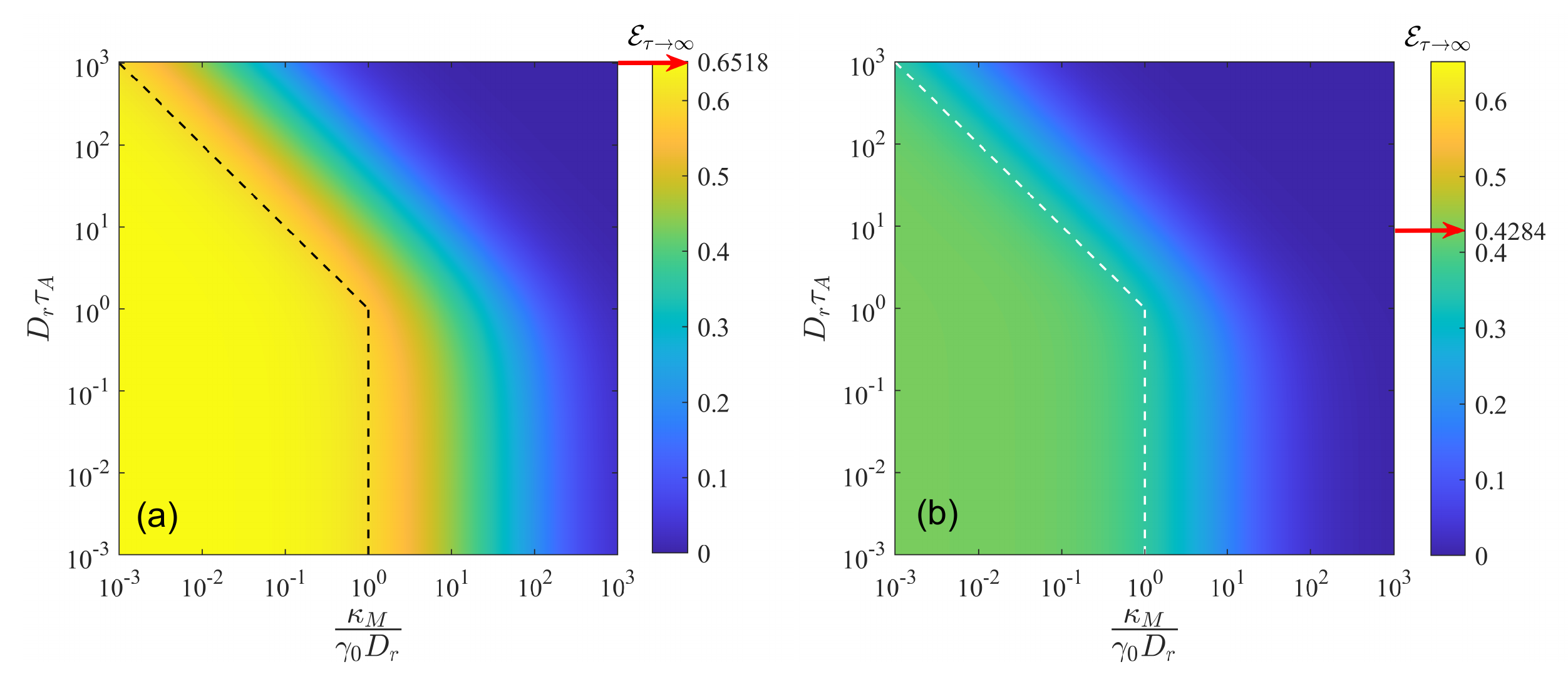}
\caption{Quasi-static efficiency of the Brownian heat engine in an active suspensions made of ABPs with non-zero persistence time, $D_r^{-1} > 0$ in a viscous solvent, under a Stirling-like cycle with $\vartheta = \frac{\kappa_M}{\kappa_m} = 10$, as a function of the normalized viscous relaxation rate $\nu_M = \frac{\kappa_M}{\gamma_0 D_r}$ and the normalized fluid relaxation time $\epsilon = D_r \tau_A$, for: (a) $\alpha = 0.1$ and $\delta = 0.1$, and (b) $\alpha = 0.9$ and $\delta = 0.9$. The red arrows in both contour plots depict the corresponding maximum efficiency given by Eq.~(\ref{eq:maximumefficiency}) under fully memoryless conditions ($\epsilon \rightarrow 0$ and $\nu_M  \rightarrow 0$). The dashed lines depict the curve given by Eq. (\ref{eq:boundary}) that separates the regions in the $\nu_M-\epsilon$ plane where the engine effectively behaves as operating at maximum quasi-static efficiency in an inert viscous bath at two effective temperatures, $T$ and $(1+\delta^{-1}) T$, (left side) or in a viscoelastic active suspension in which memory effects hinder its performance (right side).}\label{fig:4}
\end{figure*}

We now apply the generalized Langevin model (\ref{eq:GLEactive}) to investigate the performance of a Brownian engine under Stirling-like cycles by means of temporal variations of the suspension activity and the trap stiffness of the harmonic potential under isothermal conditions. For this, similar to Stirling cycles of duration $\tau >0$ imposed by temporal changes of the bath temperature, we consider the cycle represented in Figure~\ref{fig:3}, where the trap stiffness, $\kappa$, and the propulsion speed of the ABPs in the suspension, $v$, are varied in time $t$ according to the following protocols
\begin{equation}\label{eq:Stirlingkappa}
    \kappa(t) = \left\{
    \begin{array}{ll}
    \kappa_M - \frac{2}{\tau} \delta \kappa ~  t, & \,\,\,\,\, 0 \le t \le \frac{\tau}{2},\\
    \kappa_m - \delta \kappa \left(1-\frac{2}{\tau}t \right), & \,\,\,\,\, \frac{\tau}{2} < t \le \tau,
    \end{array} \right.
\end{equation}
and
\begin{equation}\label{eq:Stirlingspeed}
    v(t) = \left\{
    \begin{array}{ll}
    v_0, & \,\,\,\,\, 0 \le t < \frac{\tau}{2},\\
    0, & \,\,\,\,\, \frac{\tau}{2} \le t < \tau,\\
   v_0, & \,\,\,\,\,  t = \tau,
    \end{array} \right.
\end{equation}
respectively, where $\delta \kappa = \kappa_M - \kappa_m > 0$ and  $v_0 > 0$. During a full cycle, the actual temperature $T$ of the bath is kept constant. More specifically, a full cycle consists of a sequence of four steps:
\begin{itemize}
	\item[$1\rightarrow 2$:] {For $0 \le t < \tau/2$, the colloidal engine undergoes an expansion at constant $v(t) = v_0 $ by linearly decreasing the trap stiffness from $\kappa_M$ to $\kappa_m$}.
	\item[$2\rightarrow 3$:]{At $t = \tau/2$, the bath activity is suddenly switched off by setting $v(t) = 0$, while keeping the trap stiffness at $\kappa(t =\tau/2) = \kappa_m$, thus corresponding to a isochoric-like process.}
	\item[$3\rightarrow 4$:]{For $\tau/2 < t < \tau$, the engine undergoes a compression at zero-activity of the suspension, $v(t) = 0$, by linearly increasing the trap stiffness from $\kappa_m$ to $\kappa_M$}.
	\item[$4\rightarrow 1$:]{At $t = \tau$, the suspension activity is suddenly switched on by setting $v(t) = v_0$, while keeping the trap stiffness at $\kappa(t = \tau) = \kappa_M$, i.e. an isochoric-like process, thus  completing the full cycle.}
\end{itemize}
Then, the cycle is repeated until the system reaches a time-periodic steady state, which becomes independent of the choice of the initial condition $X(t=0)$.  

We can determine the quasi-static efficiency of this Stirling Brownian engine that would be ideally achieved in the limit $\tau \rightarrow \infty$. To this end, we first compute the mean work done by the engine during a full cycle as well as the mean heat absorbed from the environment during the active stage of the suspension. We note that, according to stochastic thermodynamics~\cite{Seifert12}, the mean work done by the system between states i and f is given by
\begin{equation}\label{eq:meanwork}
	-\langle W_{\mathrm{i} \rightarrow \mathrm{f}} \rangle = - \frac{1}{2} \int_{\mathrm{i}}^{\mathrm{f}}dt \frac{d\kappa(t)}{dt} \, \sigma^2(t),
\end{equation}
where we have expressed the fact that the variance of the particle position and the trap stiffness depend on time $t$ along the cycle, whereas the mean absorbed heat is
\begin{equation}\label{eq:meanheat}
	-\langle Q_{\mathrm{i} \rightarrow \mathrm{f}} \rangle = \frac{1}{2}\left[ \kappa_{\mathrm{f}}\sigma(\kappa_{\mathrm{f}})^2 - \kappa_{\mathrm{i}}\sigma(\kappa_{\mathrm{i}})^2 \right] - \frac{1}{2} \int_{\mathrm{i}}^{\mathrm{f}}dt \frac{d\kappa(t)}{dt} \, \sigma^2(t),
\end{equation}
with $\kappa_{\mathrm{i}}$ and $\kappa_{\mathrm{f}}$ the values of the trap stiffness in states i and f, respectively. 
In Eq. (\ref{eq:meanheat}), we have used the first law of stochastic thermodynamics, $\Delta U_{\mathrm{i} \rightarrow \mathrm{f}} = W_{\mathrm{i} \rightarrow \mathrm{f}} - Q_{\mathrm{i} \rightarrow \mathrm{f}}$, for the variation of the potential energy of the system between states i and f, where $W_{\mathrm{i} \rightarrow \mathrm{f}}$ and $Q_{\mathrm{i} \rightarrow \mathrm{f}}$ represent the corresponding \emph{work done on the system} and the \emph{heat released by the system}, respectively. We notice that, when the Stirling cycle is performed infinitely slowly, the variance of the particle position has enough time to attain the non-equilibrium steady-state value $\sigma^2$ given by Eq. (\ref{eq:varianceposition}), which is a function of $\kappa$. Therefore, by substituting such an explicit dependence of $\sigma^2$ on $\kappa$ into Eq. (\ref{eq:meanwork}), we find the mean quasi-static work done by the engine during a full Stirling-like cycle
\begin{widetext}
\begin{eqnarray}
	-\langle W_{\tau \rightarrow \infty} \rangle & = & -\langle W_{1 \rightarrow 2}\rangle - \langle W_{2 \rightarrow 3}\rangle - \langle W_{3 \rightarrow 4}\rangle  - \langle W_{4 \rightarrow 1}\rangle, \nonumber\\
						& = &  -\langle W_{1 \rightarrow 2}\rangle - \langle W_{3 \rightarrow 4}\rangle, \nonumber\\
						& = & \frac{\gamma_A(1-\alpha)v_0^2}{6 D_r} \ln \left[ \frac{\kappa_M}{\kappa_m} \left( \frac{\kappa_M+\frac{\gamma_0}{\tau_A}}{\kappa_m +\frac{\gamma_0}{\tau_A}}\right)^{\frac{\alpha \epsilon^2}{1-\alpha \epsilon^2}} \left( \frac{\kappa_m + \frac{1+{\alpha \epsilon} }{1+\epsilon}\gamma_0 D_r}{\kappa_M + \frac{1+{\alpha \epsilon} }{1+\epsilon}\gamma_0 D_r}\right)^{\frac{1}{1-\alpha \epsilon^2}}\right]. \label{eq:work}
\end{eqnarray}
\end{widetext}
Note that in above equation the isochoric works namely $\langle W_{2 \rightarrow 3}\rangle$ and $\langle W_{4 \rightarrow 1}\rangle$ cancel each other only in the quasi-static case. Likewise, from Eq.~(\ref{eq:meanheat}), we find that the mean heat quasi-statically absorbed by the engine during the first half of the cycle when the bath is active, is given explicitly by
\begin{widetext}
\begin{eqnarray}
	-\langle Q_{\tau/2 \rightarrow \infty} \rangle & = & -\langle Q_{1 \rightarrow 2} \rangle  -\langle Q_{4 \rightarrow 1}\rangle \nonumber\\
						& = & \frac{k_B T}{2} \ln \left( \frac{\kappa_M}{\kappa_m} \right) +\frac{\gamma_A(1-\alpha)v_0^2}{6  D_r} \ln \left[ \frac{\kappa_M}{\kappa_m} \left( \frac{\kappa_M+\frac{\gamma_0}{\tau_A}}{\kappa_m +\frac{\gamma_0}{\tau_A}}\right)^{\frac{\alpha \epsilon^2}{1-\alpha \epsilon^2}} \left( \frac{\kappa_m + \frac{1+\alpha \epsilon}{1+\epsilon}\gamma_0 D_r}{\kappa_M + \frac{1+\alpha \epsilon }{1+\epsilon}\gamma_0 D_r}\right)^{\frac{1}{1-\alpha \epsilon^2}}\right] \nonumber\\
						& & + \frac{\gamma_A (1-\alpha)v_0^2}{6 D_r} \left\{ 1 + \frac{1}{1-\alpha \epsilon^2} \left[  \frac{\alpha \epsilon^2 \kappa_m}{\kappa_m + \frac{\gamma_0}{\tau_A}} - \frac{\kappa_m}{\kappa_m + \frac{1+\alpha \epsilon}{1+\epsilon}\gamma_0 D_r }\right]\right\}.\label{eq:heat}
\end{eqnarray}
\end{widetext}
From Eqs.~(\ref{eq:work}) and (\ref{eq:heat}), we obtain a general expression for the efficiency of the Stirling Brownian engine operating in the viscoelastic suspension of ABPs in the quasi-static limit $\tau \rightarrow \infty$
\begin{widetext}
\begin{eqnarray}
	\mathcal{E}_{\tau \rightarrow \infty} & = & \frac{-\langle W_{\tau \rightarrow \infty} \rangle}{-\langle Q_{\tau/2 \rightarrow \infty} \rangle},\nonumber\\
								& = & \frac{\ln\left[\frac{\kappa_M}{\kappa_m}  \left( \frac{1+\frac{\kappa_M}{\gamma_0} \tau_A}{1 + \frac{\kappa_m}{\gamma_0} \tau_A} \right)^{\frac{\alpha \epsilon^2}{1-\alpha \epsilon^2}} \left( \frac{\frac{\kappa_m}{\gamma_0 D_r} + \frac{1+\alpha \epsilon}{1+\epsilon}}{\frac{\kappa_M}{\gamma_0 D_r} + \frac{1+\alpha \epsilon}{1+\epsilon}} \right)^{\frac{1}{1-\alpha \epsilon^2}}  \right]}{1+\frac{1}{1-\alpha \epsilon^2}\left(\frac{\alpha\epsilon^2\frac{\kappa_m}{\gamma_0}\tau_A}{1+ \frac{\kappa_m}{\gamma_0}\tau_A} - \frac{\frac{\kappa_m}{\gamma_0 D_r}}{\frac{\kappa_m}{\gamma_0 D_r} + \frac{1+\alpha \epsilon}{1+\epsilon}}   \right) + \ln\left[\left(\frac{\kappa_M}{\kappa_m} \right)^{1+\delta} \left( \frac{1+\frac{\kappa_M}{\gamma_0} \tau_A}{1 + \frac{\kappa_m}{\gamma_0} \tau_A} \right)^{\frac{\alpha \epsilon^2}{1-\alpha \epsilon^2}} \left( \frac{\frac{\kappa_m}{\gamma_0 D_r} + \frac{1+\alpha \epsilon}{1+\epsilon}}{\frac{\kappa_M}{\gamma_0 D_r} + \frac{1+\alpha \epsilon}{1+\epsilon}} \right)^{\frac{1}{1-\alpha \epsilon^2}}  \right]},\label{eq:efficiencyactive}
\end{eqnarray}
\end{widetext}
where we have introduced the dimensionless parameter
\begin{equation}\label{eq:delta}
	\delta = \frac{3 k_B T D_r}{\gamma_A \left(1 - \alpha\right) v_0^2} > 0,
\end{equation}
that weighs the importance of the thermal fluctuations in the system with respect to the active fluctuations induced by the self-propelled particles in the suspension. 
Some relevant limits of the quasi-static efficiency given by Eq.~(\ref{eq:efficiencyactive}) are the following:
\begin{itemize}
	\item{For $\tau_A \rightarrow 0$, $0<D_r <+ \infty$, $0 < \alpha < 1$: $\epsilon \rightarrow 0$ (memoryless friction kernel)
\begin{equation}\label{eq:memorylessefficiency}
	\mathcal{E}_{\tau \rightarrow \infty} = \frac{\ln \left[ \frac{\kappa_M}{\kappa_m} \left( \frac{1+\frac{\kappa_m}{\gamma_0 D_r}}{1+\frac{\kappa_M}{\gamma_0 D_r}} \right) \right] }{\ln \left[ \left( \frac{\kappa_M}{\kappa_m} \right)^{1+\delta} \left( \frac{1+\frac{\kappa_m}{\gamma_0 D_r}}{1+\frac{\kappa_M}{\gamma_0 D_r}} \right) \right]  + \frac{1}{1+\frac{\kappa_m}{\gamma_0 D_r}}},
\end{equation}
which is in full agreement with the expression derived in \cite{zakine2017} for an instantaneous friction kernel, \emph{i.e.}, $\Gamma(t-s) = 2 \gamma_0 \delta(t-s)$, $t \ge 0$, and $\delta = 0$ (no thermal fluctuations). Note that the prefactor $\tau_r^{-1} = D_r$ in the autocorrelation function of the active Ornstein-Uhlenbeck noise with persistence time $\tau_r = D_r^{-1}$ studied in \cite{zakine2017}, which gives rise to a white noise in the limit $\tau_r \rightarrow 0$, is absorbed by the exponent $\delta$ in Eq.~(\ref{eq:memorylessefficiency}). This suggests that such an inverse dependence on the persistence time has no effect on the quasi-static performance of the Stirling engine. Indeed, a straightforward calculation of the quasi-static efficiency for the model (\ref{eq:modifiedLangevin}) using an active Ornstein-Uhlenbeck noise with correlation $\langle \xi_A(t) \xi_A(t') \rangle = \frac{\gamma_0 k_B T_A} {\tau_r} \exp \left(- \frac{|t-t'|}{\tau_r}\right)$, where $T_A > 0$ represents an active temperature related to the strength of the active fluctuations, leads to the same Eq.~(\ref{eq:memorylessefficiency}), with a parameter $\delta$ given by $\delta = T / T_A$, which simply quantifies the relative effect of the active noise with respect to thermal fluctuations.
}
\item{For $D_r \rightarrow \infty$, $0 < \tau_A <+ \infty$, $0 < \alpha < 1$: $\epsilon \rightarrow +\infty$ (vanishingly short rotational diffusion time of the ABPs)
\begin{eqnarray}
	\mathcal{E}_{\tau \rightarrow \infty} & = & \frac{\ln \left[ \frac{\kappa_M}{\kappa_m} \left( \frac{1+\frac{\kappa_m}{\gamma_0}\tau_A}{1+\frac{\kappa_M}{\gamma_0}\tau_A} \right) \right] }{\ln \left[ \left( \frac{\kappa_M}{\kappa_m} \right)^{1+\delta} \left( \frac{1+\frac{\kappa_m}{\gamma_0}\tau_A}{1+\frac{\kappa_M}{\gamma_0}\tau_A} \right) \right]  + \frac{1}{1+\frac{\kappa_m}{\gamma_0}\tau_A}},\nonumber\\
		& \rightarrow & 0,
\end{eqnarray}
where we have used the fact that $\delta \rightarrow \infty$ when $D_r \rightarrow \infty$. This result can be explained by the fact that a vanishing rotational diffusion time of the active particles correspond to a situation in which they exhibit a negligible persistence length $D_r^{-1}v_0$ even if $v_0 >0$, thus effectively behaving as passive Brownian particles. Therefore, in this case the suspension can be regarded as a passive viscoelastic fluid from which no net work can be produced on average at constant $T$. }
\item{For $\tau_A \rightarrow \infty$, $0 < D_r <+ \infty$, $0 < \alpha < 1$: $\epsilon \rightarrow +\infty$ (infinitely large viscoelastic relaxation time of the active suspension)
\begin{equation}
	\mathcal{E}_{\tau \rightarrow \infty}  \rightarrow  \frac{\ln \left[ \frac{\kappa_M}{\kappa_m}  \frac{\frac{\kappa_m}{\gamma_0}\tau_A}{\frac{\kappa_M}{\gamma_0}\tau_A}  \right] }{\ln \left[ \left( \frac{\kappa_M}{\kappa_m} \right)^{1+\delta}  \frac{\frac{\kappa_m}{\gamma_0}\tau_A}{\frac{\kappa_M}{\gamma_0}\tau_A}  \right]} =  0,
\end{equation}
which represents the situation in which the ABPs are not able to mechanically respond in a finite time to the interaction with the PBP, hence no net work can be done on average by the engine. }
\item{For $\alpha \rightarrow 0$, $0 < \tau_A <+ \infty$, $0 < D_r < +\infty$ (very crowded active environment or very large friction coefficient of the ABPs)
\begin{equation}\label{eq:elasticefficiency}
	\mathcal{E}_{\tau \rightarrow \infty} = \frac{\ln \left[ \frac{\kappa_M}{\kappa_m} \frac{1+(1+\epsilon)\frac{\kappa_m}{\gamma_0 D_r}}{1+(1+\epsilon)\frac{\kappa_M}{\gamma_0 D_r}} \right] }{\ln \left[ \left( \frac{\kappa_M}{\kappa_m} \right)^{1+\delta} \frac{1+(1+\epsilon)\frac{\kappa_m}{\gamma_0 D_r}}{1+(1+\epsilon)\frac{\kappa_M}{\gamma_0 D_r}} \right]  + \frac{1}{1+(1+\epsilon)\frac{\kappa_m}{\gamma_0 D_r}}},
\end{equation}
which has the same form as in \cite{zakine2017} for the memoryless case, but with an effective trap stiffness $(1+\epsilon)\kappa > \kappa$, which is enhanced because of the crowded elastic surroundings.}
\item{For $\alpha \rightarrow 1$, $0 < \tau_A <+ \infty$, $0 < D_r < +\infty$: $\delta \rightarrow \infty$ (very dilute active environment or very small friction coefficient of the ABPs)
\begin{equation}\label{eq:diluteefficiency}
	\mathcal{E}_{\tau \rightarrow \infty} \rightarrow 0,
\end{equation}
which results from the fact that, in this case the thermal noise dominates over the active one, thereby leading to a vanishing mean work produced during the cycle at constant $T$.}
\end{itemize}


By defining the additional dimensionless parameters
\begin{equation}\label{eq:vartheta}
	\vartheta = \frac{\kappa_M}{\kappa_m},
\end{equation}
\emph{i.e.}, $1< \vartheta < \infty$, and
\begin{equation}\label{eq:nuM}
	\nu_M = \frac{\kappa_M}{\gamma_0 D_r},
\end{equation}
where we rule out the unphysical case $D_r = 0$, which would correspond to active particles with infinite persistence, then the expression (\ref{eq:efficiencyactive}) for the quasi-static efficiency can be written as
\begin{widetext}
\begin{equation}\label{eq:quasistatefficiency}
	\mathcal{E}_{\tau \rightarrow \infty} = \frac{\ln \left[ \vartheta \left( \frac{1+\nu_M \epsilon}{1+\frac{\nu_M}{\vartheta}\epsilon} \right)^{\frac{\alpha \epsilon^2}{1-\alpha \epsilon^2}} \left( \frac{\frac{\nu_M}{\vartheta}+\frac{1+\alpha \epsilon}{1+\epsilon}}{\nu_M+\frac{1+\alpha \epsilon}{1+\epsilon}} \right)^{\frac{1}{1-\alpha \epsilon^2}}\right]}{1 + \frac{1}{1-\alpha \epsilon^2} \left( \frac{\alpha \nu_M \epsilon^3}{\vartheta+\nu_M\epsilon} - \frac{\nu_M}{\nu_M+\frac{1+\alpha \epsilon}{1+\epsilon}\vartheta} \right) + \ln \left[  \vartheta^{1+\delta} \left( \frac{1+\nu_M \epsilon}{1+\frac{\nu_M}{\vartheta}\epsilon} \right)^{\frac{\alpha \epsilon^2}{1-\alpha \epsilon^2}} \left( \frac{\frac{\nu_M}{\vartheta}+\frac{1+\alpha \epsilon}{1+\epsilon}}{\nu_M+\frac{1+\alpha \epsilon}{1+\epsilon}} \right)^{\frac{1}{1-\alpha \epsilon^2}} \right]}.
\end{equation}
\end{widetext}
In Fig. \ref{fig:4} we represent as contour plots the behavior of the quasi-static efficiency, described by Eq. (\ref{eq:quasistatefficiency}), as a function of $\nu_M$ and $\epsilon$ for a fixed value of the ratio of the maximum and minimum trapping constants  ($\vartheta = 10$) and for two representative values of the ratio of friction coefficients: $\alpha = 0.1$ and $\alpha = 0.9$.  In both cases, we set the value of the ratio $\frac{k_B T D_r}{\gamma_A v_0^2}$ to 0.03, in such a way that the parameter $\delta$ defined in Eq.~(\ref{eq:delta}) has the values $\delta = 0.1$ for $\alpha = 0.1$, whereas $\delta = 0.9$ for $\alpha = 0.9$. As can be seen in both contour plots, for a given value of $\epsilon \lesssim 1$, the profile of $\mathcal{E}_{\tau \rightarrow \infty}$ as a function of $\nu_M$ remains rather unaltered by $\epsilon$, thus being very similar to the dependence on $\nu_M$ for the memoryless case described by Eq. (\ref{eq:memorylessefficiency}). This dependence exhibits two distinctive features. For sufficiently small $\nu_M$, typically $\nu_M \lesssim 1$, $\mathcal{E}_{\tau \rightarrow \infty}$ is comparatively constant, while for $\nu_M \gtrsim 1$, an evident monotonic decrease of the efficiency is observed with increasing values of $\nu_M$, thereby becoming zero at a certain value of $\nu_M$ that depends on the specific values of $\alpha$ and $\vartheta$. However, if $\epsilon \gtrsim 1$, there is a qualitative change in the dependence of $\mathcal{E}_{\tau \rightarrow \infty}$ on $\nu_M$, where the monotonic decrease to the operation regime with zero quasi-static-efficiency  occurs at smaller and smaller values of $\nu_M$ with increasing $\epsilon$. In Figures \ref{fig:4}(a) and (b), for comparison we also show as arrows the corresponding values of the quasi-static efficiency that would be achieved under fully memoryless conditions, \emph{i.e.}, $\epsilon \rightarrow 0$ and $\nu_M  \rightarrow 0$
\begin{equation}\label{eq:maximumefficiency}
	\mathcal{E}_{\mathrm{max}} = \frac{1}{1+\delta +\frac{1}{\ln \vartheta}},
\end{equation}
thereby verifying that Eq. (\ref{eq:maximumefficiency}) establishes an upper limit on the efficiency of the Brownian engine operating under the Stirling-like cycle described by Eqs. (\ref{eq:Stirlingkappa}) and (\ref{eq:Stirlingspeed})
\begin{equation}\label{eq:maxeff}
	\mathcal{E}_{\tau \rightarrow \infty}  \le \mathcal{E}_{\mathrm{max}} 
\end{equation}
regardless of the values of $\nu_M$ and $\epsilon$ that characterize the coupling with the active suspension. Indeed, since $\vartheta \ge 1$ by definition, from Eq. (\ref{eq:quasistatefficiency}) it can be readily demonstrated that, if $\epsilon$ and $\nu_M$ are simultaneously much smaller than 1, or $\nu_M \ll \epsilon^{-1}$ if $\epsilon \gtrsim 1$, then the quasi-static efficiency of the Stirling engine converges to that given by Eq. (\ref{eq:maximumefficiency}). Therefore, the curve
\begin{equation}\label{eq:boundary}
    \nu_M = \left\{
    \begin{array}{ll}
    1, & \,\,\,\,\, 0 \le \epsilon \le 1,\\
   \epsilon^{-1}, & \,\,\,\,\,  \epsilon > 1,
    \end{array} \right.
\end{equation}
defines two distinct regimes in the operation of the Brownian engine corresponding to two distinct regions in the in the $\nu_M-\epsilon$ plane of the system, as depicted by the two regions separated by the dashed lines in Figs. \ref{fig:4}(a) and (b). The first regime corresponds to a situation in which the engine operates very close to its maximum possible efficiency given by Eq. (\ref{eq:maximumefficiency}), see yellow and green regions on the left sides of Figs. \ref{fig:4}(a) and (b), respectively. This can be regarded as a regime in which the quasi-static efficiency is equivalent to that of a Brownian Stirling engine under a periodic variation of the actual temperature of an inert (passive) viscous bath alternating between the values $T$ and $(1+\delta^{-1}) T$, where the corresponding Carnot efficiency 
\begin{equation}\label{eq:Carnot}
\mathcal{E}_C = 1- \frac{T}{\left(1+\frac{1}{\delta}\right)T} = \frac{1}{1+\delta},
\end{equation}
can only be achieved if $\vartheta \rightarrow \infty$, in such a way that $\mathcal{E}_{\tau_\rightarrow \infty} \approx \mathcal{E}_{\mathrm{max}} \le \mathcal{E}_C$ for any finite value of $\vartheta$. On the other hand, the second regime of operation correspond to a situation in which memory effects originating either from the viscoelasticity of the active suspension or from the persistence in the self-propulsion of the ABPs are sufficiently strong to hinder the operation of the engine. This leads to an underperformance with respect to the maximum efficiency $\mathcal{E}_{\mathrm{max}}$, as shown in the regions to the right of the dashed lines in Figs. \ref{fig:4}(a) and (b). We note that a similar expression for efficiency as Eq. (\ref{eq:maximumefficiency}) was also derived in some recent works \cite{Marathe22, Lee2021}.

\begin{figure}
    \centering
\includegraphics[width=1\columnwidth]{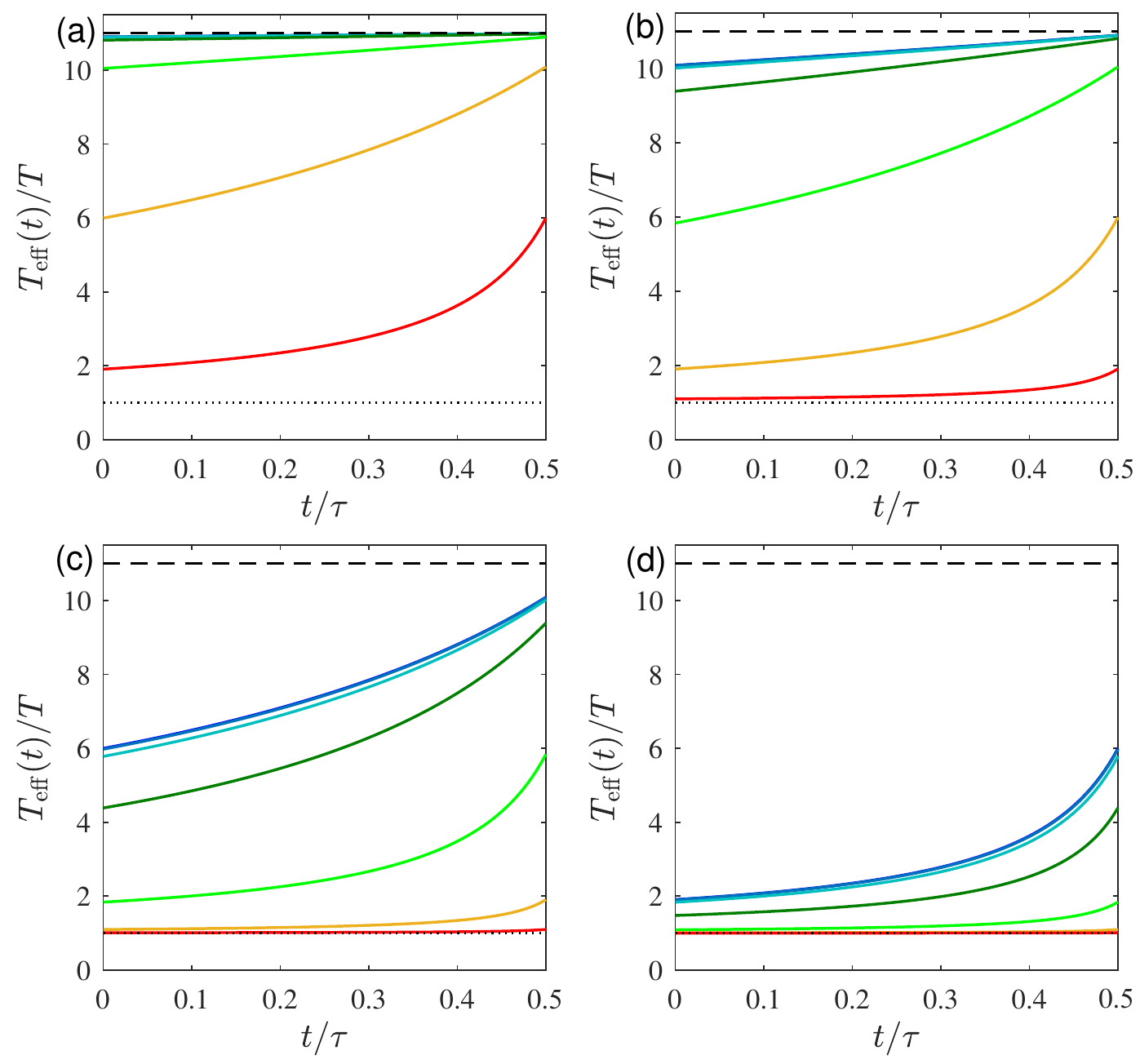}
\caption{Quasi-static effective temperature of the PBP, defined in Eq. (\ref{eq:effectivetemperature}), as a function of time (normalized by the cycle period $\tau$) during the active stage of a Stirling-like cycle ($0 \le t < \tau /2$) with $\vartheta = 10$, $\delta = 0.1$, $\alpha = 0.1$, and distinct values of the parameter $\nu_M = \frac{\kappa_M}{\gamma_0 D_r}$: (a)  $\nu_M = 0.01$, (b)  $\nu_M = 0.1$, (c)  $\nu_M = 1$, and (d)  $\nu_M = 10$. In all subfigures, the solid lines in different colors represent distinct values of the parameter $\epsilon = D_r \tau_A$. From top to bottom: $\epsilon = 0.001$ (dark blue), $\epsilon = 0.01$ (light blue), $\epsilon = 0.1$ (turquoise), $\epsilon = 1$ (dark green), $\epsilon = 10$ (light green), $\epsilon = 100$ (orange), $\epsilon = 1000$ (red), whereas the black dashed and dotted lines represent $T_{\mathrm{eff}}/T = 1 + \delta^{-1} = 11$ and $T_{\mathrm{eff}}/T = 1$, which correspond to the maximum and minimum possible values of the effective temperature that can be reached by the interaction of the PBP with the active suspension.} \label{fig:5}
\end{figure}

At first glance, the previous results seem to be contrary to some recent studies \cite{Marathe22, Lee2021}, where it was shown that non-Gaussianity is not important but non-Markovianity of the bath is essential to improve the performance of a cyclic active engine over its passive counterpart. However, it should be noted that, in both of these studies, as well as in most investigations of microscopic engines in active baths, the non-Markovianity of the system is only due to non-trivial correlations of the active noise, which are directly related to the persistence time of the active particles in the bath, whereas friction forces are memory-less \cite{zakine2017, Arnab18, saha2019, Arnab20, Steffenoni20, Viktor20, Fodor20, Slahiri21}. On the other hand, in the present work, apart from the persistence time, which we denote as $D_r^{-1}$, we also take into account the non-Markovianity of the system due to the slow mechanical relaxation of the surrounding ABPs interacting with the PBP, both emerging naturally from the coarse-grained description of the PBP dynamics. This results in memory friction with a characteristic viscoelastic relaxation time $\tau_A$ that is of different physical origin from $D_r^{-1}$, which also affects the active noise correlations, as described by Eq. (\ref{eq:activenoise}). We point out that the non-Markovianity considered in previous works (finite $D_r^{-1} > 0$ and $\tau_A = 0$) actually corresponds to the case $\epsilon = 0$ [see Eq. (\ref{eq:memorylessefficiency})], which agrees with the efficiency derived in \cite{zakine2017}. In the present case, motivated by the concept of effective temperature in active engines \cite{zakine2017,Viktor20,Steffenoni20}, we can define an effective temperature in the quasi-static limit through the variance $\sigma(t)^2$ of the PBP position within the harmonic trap, see Eq. (\ref{eq:varianceposition})
\begin{widetext}
\begin{eqnarray}\label{eq:effectivetemperature}
    T_{\mathrm{eff}}(t) & = & \frac{\kappa}{k_B} \sigma(t)^2\nonumber\\
    & = &  \left\{ 1 + \frac{1+[\alpha+\nu(t)]\epsilon}{[1+\nu(t) \epsilon] \left\{ 1 + \nu(t) + [\alpha + \nu(t)]\epsilon \right\} } \frac{1}{\delta(t)} \right\} T.
\end{eqnarray}
\end{widetext}
The effective temperature of the engine defined in Eq. (\ref{eq:effectivetemperature}) depends on the instantaneous values at time $t$ of the trap stiffness $\kappa(t)$ and the ABPs self-propulsion speed $v(t)$  through the time-dependent parameters $\nu(t) = \frac{\kappa(t)}{\gamma_0 D_r}$ and $\delta(t) = \frac{3 k_B T D_r}{\gamma_A (1-\alpha) v(t)^2}$, respectively, see Eqs. (\ref{eq:nu}) and (\ref{eq:delta}). Consequently, such an effective temperature is in general time-dependent along the cycle and reduces to the bath temperature $T_{\mathrm{eff}}(t) = T$ during the passive stage of the suspension ($\tau/2 \le t < \tau$) where $v(t) = 0$ and $\delta(t) \rightarrow \infty$. In Fig. \ref{fig:5} we illustrate the typical time evolution of $T_{\mathrm{eff}}(t)$ during the active half of the Stirling cycle (\ref{eq:Stirlingspeed}) ($0 \le t < \tau/2$) with $\vartheta = 10$, $\delta = 0.1$, $\alpha = 0.1$ and distinct values of $\epsilon$ spanning six orders of magnitude ($\epsilon = 10^{-3}-10^3$) for four exemplary values of $\nu_M$ ($0.01,0,1,1,10$). From these plots, it evident that at sufficiently small values of $\nu_M$ and $\epsilon$, the effective temperature experienced by the PBP during the active stage of the Stirling cycle remains rather close to the maximum possible value $(1+\delta^{-1})T$ corresponding to the fully Markovian case $\epsilon = \nu_M = 0$, see Figs. \ref{fig:5}(a) and (b) for $\nu_M = 0.01, 0.1$ and $\epsilon < 1$. Nevertheless, as $\nu_M$ or $\epsilon$ increase, the values of the effective temperature are in general well below $(1+\delta^{-1})T$, as can be seen in Fig. \ref{fig:5}(c) for $\nu_M = 1$. In the extreme case where $\nu_M \gg 1$ and $\epsilon \gg 1$, $T_{\mathrm{eff}}(t)$ approaches the value $T$, \emph{i.e.}, the bath temperature of the solvent in which the ABPs and the PBP are immersed, as verified in Fig. \ref{fig:5}(d) for $\nu_M = 10$ and $\epsilon = 10, 100, 1000$. Therefore, when non-Markovianity due to the ABPs persistent motion and the viscoelastic behavior of the active suspension are both dominant over viscous dissipation of the PBP within the harmonic trap, it is expected that the efficiency of the Stirling-like engine is highly reduced. Indeed, from Eq. (\ref{eq:effectivetemperature}), we can clearly see that in the limit $\epsilon \rightarrow 0$ and $\nu(t) \rightarrow 0$, \emph{i.e.}, $\tau_A \ll D_r^{-1} \ll \frac{\gamma_0}{\kappa(t)}$, \emph{i.e.} when the active suspension behaves as a memoryless active bath with instantaneous friction and small persistence time, the effective temperature of the engine during the active stage of the cycle converges to the value $(1 + \delta^{-1}) T$, at which the maximum efficiency given by Eq. (\ref{eq:maximumefficiency}) is achieved.  If this condition on the time-scales is not fulfilled, memory effects tend to reduce the effective temperature of the system and consequently also the efficiency of the engine. 

The physical origin of such a nontrivial behavior of $T_{\mathrm{eff}}(t)$ can be traced back to the decrease in the variance $\sigma_A^2$ of the PBP position with increasing values of $\epsilon$, as seen in Fig. \ref{fig:2}, In turn, this is a consequence of the interaction with the ABPs in the suspension, which results in three effective forces acting on the PBP: a frictional force, a fluctuating force of thermal original and a stochastic force associated to the self-propulsion of the ABPs, being the first two closely related by Eq. (\ref{eq:thermalnoise}). If the mechanical relaxation of the active suspension is much faster than the rotational diffusion time of the ABPs ($\tau_A \ll D_r^{-1}$, memory-less friction), then the variance $\sigma^2$ is independent of $\tau_A$. However, if $\tau_A \gtrsim D_r^{-1}$, there is a reduction in the high frequency components of the power spectrum of the active fluctuations with respect to the case of instantaneous friction, see Fig. \ref{fig:1}(c), because the ABPs are not able to respond as quickly as in the latter case. For a fixed value of the trap stiffness of the harmonic potential, this leads to a reduction in the active component of the variance of the PBP position and a decrease of the corresponding effective temperature.

\begin{figure*}
    \centering
\includegraphics[width=2\columnwidth]{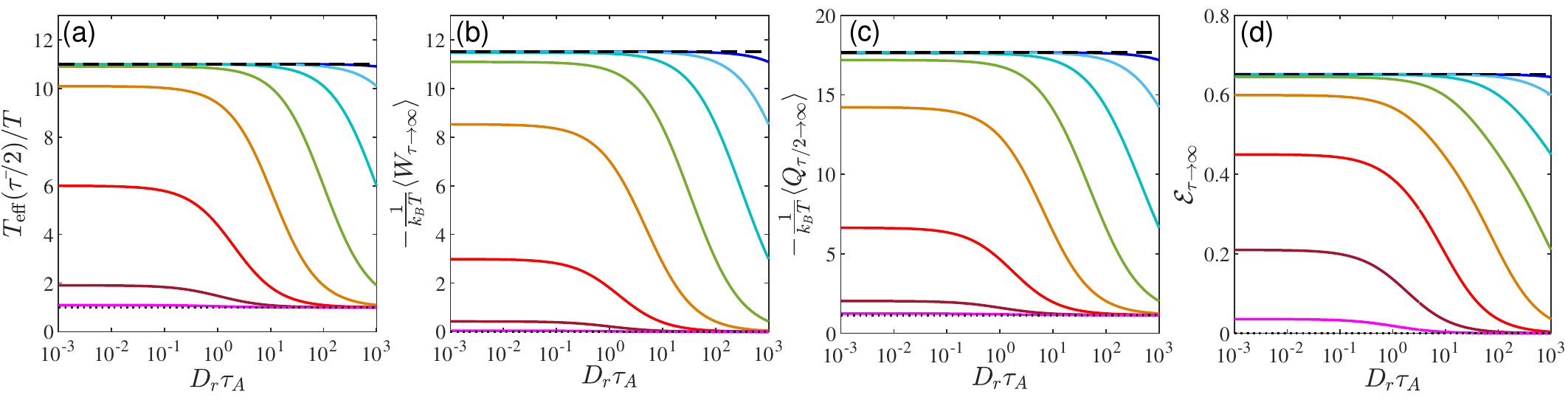}
\caption{Dependence on the parameter $\epsilon = D_r \tau_A$ of (a) the effective temperature defined in Eq. (\ref{eq:effectivetemperature}) at time $t = \frac{1}{2}\tau^{-}$; (b) the mean work delivered during a Stirling cycle; (c) the mean heat absorbed during the first half of the cycle; and (d) the efficiency of the engine, for $\alpha = 0.1$, $\delta = 0.1$ and $\vartheta = 10$, and different values of the ratio $\nu_M = \frac{\kappa_M}{\gamma_0 D_r}$ (solid lines). From top to bottom: $\nu_M = 10^{-4}$ (dark blue), $10^{-3}$ (light blue), $10^{-2}$ (turquoise), $10^{-1}$ (green), $10^{0}$ (orange), $10^{1}$ (red), $10^2$ (garnet red), and $10^3$ (magenta). The black dashed and dotted lines represent the limiting values of the corresponding quantities for  $\nu_M = 0$ ($T_{\mathrm{eff}} = 11 T$, $-\langle W_{\tau \rightarrow \infty} \rangle = 11.5129 k_B T$, $-\langle Q_{\tau/2 \rightarrow \infty} \rangle = 17.6642 k_B T$, $\mathcal{E}_{\tau \rightarrow \infty} = 0.6518$), and $\nu_M \rightarrow \infty$ ($T_{\mathrm{eff}} = T$, $-\langle W_{\tau \rightarrow \infty} \rangle = 0$, $-\langle Q_{\tau/2 \rightarrow \infty} \rangle = 1.1513 k_B T$, $\mathcal{E}_{\tau \rightarrow \infty} = 0$), respectively.}\label{fig:6}
\end{figure*}

In Figs.~\ref{fig:6}(a)-(d), we verify that the values of the effective temperature of the system  have a direct effect in the values of the
mean work delivered during a Stirling cycle the mean heat absorbed during the first half of the cycle, and consequently affect the efficiency of the engine in a straightforward manner. First, in Fig. \ref{fig:6}(a) we plot the values of the quasi-static effective temperature (normalized by the bath temperature $T$) at the end of the first half of the Stirling cycle ($t = \frac{1}{2}\tau^{-}$) as a function of the parameter $\epsilon$ for $\alpha = 0.1$, $\delta = 0.1$, $\vartheta = 10$, and different values of $\nu_M$. We observe that, when the effective temperature is close to its maximum value $\left(1+\frac{1}{\delta}\right)T$, the mean work and the mean absorbed heat are also close to the their maximum values  $-\langle W_{\tau \rightarrow \infty} \rangle = \frac{k_B T}{2\delta} \ln \vartheta$, $-\langle Q_{\tau/2 \rightarrow \infty} \rangle = \frac{k_B T}{2}\left( 1 + \frac{1}{\delta} \right) \ln \vartheta + \frac{k_B T}{2\delta}$, as verified in Figs. \ref{fig:6}(b) and (c), respectively. Consequently, in this case the efficiency is also close to its maximum value given by Eq. (\ref{eq:maximumefficiency}). Decreasing the effective temperature of the system due to viscoelastic and self-propulsion memory effects of the active suspension translates generally into a reduction of the mean work and the mean absorbed heat, where the former is more pronounced that the latter, thereby resulting in a decrease of the efficiency. Finally, when the effective temperature is close to the bath temperature, the mean work and the mean heat approach the values $-\langle W_{\tau \rightarrow \infty} \rangle \rightarrow 0$ and $-\langle Q_{\tau/2\rightarrow \infty} \rangle \rightarrow \frac{k_B T}{2} \ln \vartheta > 0$, respectively, which correspond to the situation $\nu_M \rightarrow \infty$. This implies that $\mathcal{E}_{\tau \rightarrow \infty } \rightarrow 0$ when memory effects due to both the viscoelasticity of the suspension and the self-propulsion of the ABPs are dominant over the viscous relaxation of the PBP in the harmonic potential.

\section{Concluding remarks}
\label{sect:conclude}
In this paper, we have studied a model for a Brownian Stirling-like engine functioning under isothermal conditions in a suspension of active particles, whose self-propulsion speed can be periodically tuned in time. By integrating out the degrees of freedom of the active particles in the suspension, we have derived a generalized Langevin equation that effectively describes the stochastic motion of an embedded passive particle trapped by a harmonic potential, which includes memory friction and persistent self-propelled motion of the active particles. Under Stirling-like cycles implemented by means of periodic variations of the trap stiffness and the propulsion speed of the active particles, the heat absorbed from the active suspension can be effectively converted into net work on average even at constant temperature of the bath. From the Langevin model, we are able to derive an analytical expression for the efficiency of the engine in the quasi-static limit of operation, which correctly reduces to that in the case of a memoryless friction kernel reported in the literature. We find that the main effect of the viscoelastic relaxation time of the active suspension is to reduce the efficiency of the Brownian Stirling engine operating in the quasi-static limit, as compared to the case of a system with instantaneous friction. We point out that, unlike earlier phenomenological models of active noise, our approach reveals two sources of non-Markovianity that naturally emerge from the interaction between the Brownian particle and its non-equilibrium environment, namely the friction kernel and active noise correlations. Thus, we attribute the reduction in efficiency to the competition between the times scales involved in these two sources. Indeed, all the thermodynamics quantities depend on the active part of the positional variance $\sigma^2_A$, which is highly reduced when the viscoelastic relaxation time of the suspension is larger than the rotational diffusion time of the active particles and the viscous relaxation time of the trapped particle. Thus, a sharp decrease in it gives rise to a large reduction in the effective temperature of the system $T_{\mathrm{eff}}$ leading to an appreciable lowering in the efficiency of the active engine. Nevertheless, there is a broad interval of values of the ratio between the viscoelastic relaxation time and the rotational diffusion time of the active particles in which such memory effects do not give rise to a reduction of the quasi-static efficiency. In such a regime, the Stirling-like cycle performed by the variation of the suspension activity behaves as an actual Stirling cycle, whose efficiency is bounded by an equivalent Carnot efficiency with a well-defined effective temperature directly related to the squared propulsion velocity of the active particles. Apart from the derivation of such a theoretical bound from first principles, our ideas could be experimentally realized in mesoscopic soft-matter systems by using tunable baths composed of, e.g., photosensitive bacteria~\cite{walter2007,arlt2018}, hot Brownian nanoparticles~\cite{huang2021,kumar2020}, chemically fueled~\cite{Ebbens2016} or light-activated colloids \cite{GomwzSolano2017,Vutukuri2020,gomez_solano2020,gomez_solano2022}, without the need for artificially generated active noise, nor cyclic temporal changes in the actual bath temperature, which are generally difficult to control in experiments. We believe that investigating the non-quasi-static functioning of such isothermal active engines as well as other regimes of operation could also be interesting for practical applications and will be studied elsewhere.

\section*{Acknowledgments}
J. R. G.-S. acknowledges support from DGAPA-UNAM PAPIIT Grant No. IA104922. R. M. gratefully acknowledges Science and Engineering Research Board (SERB), India for financial support through the MATRICS Grant (No. MTR/2020/000349).

\appendix*

\section{Derivation of a generalized Langevin equation for a confined Brownian particle in an active suspension}\label{app:derivGLE}

In this Appendix, we provide more details on the derivation of the generalized Langevin equation for a passive Brownian particle confined by a harmonic potential and surrounded by a suspension of active Brownian particles. We start from the description of the stochastic motion of the three-dimensional position of the PBP at time $t$, $\mathbf{R}(t) = [X(t),Y(t),Z(t)]$, which in the overdamped limit reads
\begin{equation}\label{eq:Langevin0}
 \gamma_{\infty}\frac{d}{dt}\mathbf{R}(t) = -\nabla_{\mathbf{R}}U(\mathbf{R}(t)) + \mathbf{G}(\left\{ \mathbf{r}_i(t) \right\},\mathbf{R}(t)) + \bm{\zeta}_{\infty}(t).
\end{equation}
In Eq. (\ref{eq:Langevin0}), $\gamma_{\infty}$ is the friction coefficient experienced by the particle in a viscous solvent of constant viscosity $\eta_{\infty}$, $U(\mathbf{R}(t)) = \frac{1}{2} \kappa |\mathbf{R}(t)|^2$ represents the an external harmonic potential that confines the particle motion,  $\mathbf{G}(\left\{ \mathbf{r}_i(t) \right\},\mathbf{R}(t))$ is the interaction force between the Brownian particle and smaller self-propelled particles, whose dynamics are described by the set of spatial coordinates $\{ \mathbf{r}_i(t) \}$, where the index $i = 1, ..., N$ labels each component of the suspension formed by a total of $N \gg 1$ active particles. Moreover, $\bm{\zeta}_{\infty}(t)$ is a Gaussian white noise which mimics the effect of the thermal collision between the solvent molecules and the trapped Brownian particle. The mean value and the autocorrelation function of $\bm{\zeta}_{\infty} (t)$ are
\begin{eqnarray}\label{eq:whitenoise1}
    \langle \bm{\zeta}_{\infty}(t) \rangle & = & \mathbf{0},\nonumber\\
    \langle \bm{\zeta}_{\infty}(t) \otimes \bm{\zeta}_{\infty}(s) \rangle & = & 2k_B T \gamma_{\infty}\delta(t-s) \mathbb{I}
\end{eqnarray}
respectively, where $T$ is the temperature of a heat bath with which the system is kept in contact, $\otimes$ stands for the dyadic product, and $\mathbb{I}$ represents the identity tensor. In addition, we assume that the interaction force derives from the potential $V(\left\{ \mathbf{r}_i(t) \right\},\mathbf{R}(t))$, \emph{i.e.}, $ \mathbf{G}(\left\{ \mathbf{r}_i(t) \right\},\mathbf{R}(t)) = - \nabla_{\mathbf{R}} V(\left\{ \mathbf{r}_i(t) \right\}, \mathbf{R}(t))$. On the other hand, the dynamics of the ABPs at time $t$ are described by the stochastic model
\begin{equation}\label{eq:ABMmodel}
    \frac{d}{dt} \mathbf{r}_{i}(t) = v \mathbf{n}_i(t) - \mu \nabla_{\mathbf{r}_i} V(\left\{ \mathbf{r}_i(t) \right\}, \mathbf{R}(t)) +  \bm{\chi}_i(t),
\end{equation}
where $v$ is the self-propulsion speed of the active particles, $\mathbf{n}_i(t)$ is a unit vector that represents the instantaneous orientation of the $i-$th particle, $\mu$ is the particle mobility, and $\bm{\chi}_i(t)$ is a Gaussian white noise with mean and autocorrelation function
\begin{eqnarray}\label{eq:whitenoise2}
    \langle \bm{\chi}_i(t) \rangle & = & \mathbf{0},\nonumber\\
    \langle \bm{\chi}_i(t) \otimes \bm{\chi}_{j}(s) \rangle & = & 2k_B T \mu \delta_{ij}\delta(t-s) \mathbb{I},
\end{eqnarray}
respectively, with $\delta_{ij}$ the Kronecker delta. The particle orientations have the following correlation functions
\begin{equation}\label{eq:persistorientation}
    \langle \mathbf{n}_i(t) \cdot \mathbf{n}_j(s) \rangle = \delta_{ij} \exp \left( -D_r |t-s|\right)
\end{equation}
which capture the effect of rotational diffusion, with diffusion coefficient $D_r$, on the persistence of the self-propulsion velocity. For the sake of simplicity, in Eqs. (\ref{eq:ABMmodel}) and (\ref{eq:persistorientation}), we assume that the active particles do not interact with each other but only with the bigger Brownian particle, which is sufficient to capture a non-trivial viscoelastic coupling with the surroundings.

We now proceed to average out the degrees of freedom of the ABPs in order to obtain an effective equation of motion for the PBP, which can be expressed as
\begin{equation}\label{eq:GLE1}
    \gamma_{\infty} \frac{d}{dt}\mathbf{R}(t) = -\kappa \mathbf{R}(t) + \langle \mathbf{G}(t)\rangle + \bm{\zeta}(t) + \bm{\zeta}_{\infty}(t)
\end{equation}
In Eq.~(\ref{eq:GLE1}), we denote $\mathbf{G}(t) \equiv \mathbf{G}(\left\{ \mathbf{r}_i(t) \right\},\mathbf{R}(t))$ for the sake of notation simplicity, whereas $\bm{\zeta}(t) = \mathbf{G}(t) - \langle \mathbf{G}(t) \rangle$ is a stochastic force due to the interaction with the active particles, whose first two moments will be determined in the limit $N \gg 1$, and the angle brackets denote the average over all the degrees of freedom of the active particles
\begin{equation}\label{eq:averageG}
    \langle \mathbf{G}(t) \rangle
    = \int d \mathbf{r}_1 \ldots \int d \mathbf{r}_N \rho(\left\{ \mathbf{r}_i \right\},t | \mathbf{R})\mathbf{G}(\left\{ \mathbf{r}_i \right\},\mathbf{R})
\end{equation}    
with $\rho(\left\{ \mathbf{r}_i \right\},t | \mathbf{R})$ the joint probability density of the active-particle coordinates at time $t$ conditioned by the specific position $\mathbf{R}$ of the Brownian particle. We also assume that the interaction between the passive Brownian particle and the active ones is sufficiently weak in such a way that $k \ll \kappa$. Then, we expand the interaction potential at linear order in $\mathbf{R}$ around the minimum of the external harmonic potential, $\mathbf{R} = \mathbf{0}$. For this, we assume that the interaction potential is harmonic, \emph{i.e.}, $V(\left\{ \mathbf{r}_i \right\}, \mathbf{R}) = \frac{1}{2} k \sum_{i=1}^N |\mathbf{R} - \mathbf{r}_i|^2$, which yields
\begin{equation}\label{eq:perturbenergy}
 V(\left\{ \mathbf{r}_i \right\}, \mathbf{R}) \approx  V(\left\{ \mathbf{r}_i \right\}, \mathbf{0}) - \mathbf{G}(\left\{ \mathbf{r}_i \right\}, \mathbf{0})\cdot \mathbf{R},
\end{equation}
where $\mathbf{G}(\left\{ \mathbf{r}_i \right\}, \mathbf{0})\cdot \mathbf{R} =  \sum_{j=1}^3 {G}_j(\left\{ \mathbf{r}_i \right\}, \mathbf{0}) R_j$ can be regarded as a linear perturbation to the energy of the unperturbed system characterized by $ \mathbf{R} =  \mathbf{0}$ due to the action of the active particles in the suspension, with $R_j$ the $j-$th component of the vector $\mathbf{R}$. Without loss of generality, we can focus on a single coordinate of the passive Brownian particle, \emph{e.g.}, $X = R_1$, for which we derive its corresponding  effective equation of motion by taking into account that $h(t) \equiv X(t)$ can be considered as a time-dependent external protocol perturbing the stationary state $X = 0$. According to Eq. (\ref{eq:perturbenergy}) the variable conjugate to the perturbation $h(t)$ with respect to the energy of the unperturbed system is $\mathcal{V}({t}) = G_X(\left\{ x_{i}(t) \right\}, X = 0)$, where $G_X$ and $x_i$ are the components of $\mathbf{G}$ and $\mathbf{r}_i$ along $X$, respectively. Therefore, the general linear response formula around non-equilibrium states in contact with a thermal bath~\cite{steffenoni2016,maes2015}
\begin{widetext}
\begin{equation}\label{eq:LRnoneq}
    \langle \mathcal{G}(t) \rangle =  \langle \mathcal{G}(t) \rangle^0 + \frac{1}{2k_B T} \int_{t_0}^t ds \, h(s) \left\{ \frac{d}{ds} \langle \mathcal{V}(s)  \mathcal{G}(t) \rangle^0 - \langle \left[\mathbb{L} \mathcal{V}(s) \right] \mathcal{G}(t) \rangle^0 \right\},
\end{equation}
\end{widetext}
can be applied to the specific observable $\mathcal{G}(t) = G_X( \left\{x_{i}(t) \right\}, X (t))$. In Eq. (\ref{eq:LRnoneq}), $\langle \ldots \rangle^0$ represents an average with respect to the active-particle coordinates in the unperturbed state, \emph{i.e.}, with density $\rho(\left\{ \mathbf{r}_i \right\}, t | \mathbf{R} = \mathbf{0})$ at time $t < t_0$ before the perturbation [see Eq. (\ref{eq:averageG})], whereas $\langle \ldots \rangle$ corresponds to an average in presence of the perturbation $h(t)$, which starts to act at time $t = t_0$. In addition, $\mathbb{L}$ denotes the backward generator of the unperturbed dynamics of the active suspension, which explicitly reads
\begin{eqnarray}\label{eq:Lgenerator}
    \mathbb{L} & = & \sum_{i=1}^N \left\{ \left[vn_{1i}  - \mu \partial_{x_i}V(\left\{x_i \right\}, X = 0) \right] \partial_{x_i} + \mu k_B T \partial_{x_i}^2  \right\},\nonumber \\
    & = & \sum_{i=1}^N \left[ \left( vn_{1i}  - \mu  k x_i\right) \partial_{x_i} + \mu k_B T \partial_{x_i}^2  \right],
\end{eqnarray}
where $n_{1i}$ represents the projection of the orientation of the $i-$th particle, $\mathbf{n}_i$, onto $X$. Note that the choice of the observable $\mathcal{G}$ is justified by the fact that the mean value of $G_X$ in presence of the perturbation, averaged over the ABPs coordinates, is directly involved in the equation of motion of $X$
\begin{equation}\label{eq:GLE2}
    \gamma_{\infty} \frac{d}{dt}X(t) = -\kappa X(t) + \langle G_X(t)\rangle + {\zeta}(t) + {\zeta}_{\infty}(t),
\end{equation}
where $\zeta$ and $\zeta_{\infty}$ are the components of $\bm{\zeta}$ and $\bm{\zeta}_{\infty}$ along $X$, respectively. 
By integrating by parts Eq. (\ref{eq:LRnoneq}), and taking the limit $t_0 \rightarrow -\infty$ in such a way that all  correlations with the initial condition vanish, we find the following expression for the mean value of $G_X(t)$
\begin{equation}\label{eq:meanGX}
    \langle G_X(t) \rangle = \Gamma_A(0)X(t) -  \int_{-\infty}^t ds \, \Gamma_A(t-s) \frac{d}{ds}X(s).
\end{equation}
In Eq. (\ref{eq:meanGX}), we have taken into account that $\langle \mathcal{G}(t) \rangle^0 =  k \sum_{i=1}^N \langle x_i \rangle^0 = 0$ by isotropy of the active suspension, and 
\begin{equation}\label{eq:kernelact}
    \Gamma_A(t - s) = \frac{1}{2k_B T}\left[ \langle \mathcal{V}(t)\mathcal{V}(s) \rangle^0 -\int_{-\infty}^s du \, \langle \left[ \mathbb{L}\mathcal{V}(u)\right] \mathcal{V}(t) \rangle^0 \right],
\end{equation}
is a memory kernel that emerges in the effective dynamics of the PBP due to the interactions with the ABPs. In Eq. (\ref{eq:kernelact}), we have considered that, in the unperturbed state characterized by $X=0$, within the angular brackets $\langle \ldots \rangle^0$ the observable $\mathcal{G}(t)$ reduces to $\mathcal{G}(t) = G_X(\{ x_i(t) \} , X(t) = 0) = \mathcal{V}(t)$. Furthermore, since the ABPs do not interact with each other and are identical, the different correlation functions involved in Eq. (\ref{eq:kernelact}) are simply given by
\begin{widetext}
\begin{equation}\label{eq:correlkernel1}
    \langle \mathcal{V}(t)\mathcal{V}(s) \rangle^0 = N k^2 \langle x_1(t) x_1(s) \rangle, 
\end{equation}
\begin{equation}\label{eq:correlkernel2}
    \langle \left[ \mathbb{L}\mathcal{V}(u)\right] \mathcal{V}(t) \rangle^0 = N k^2 \left[ v \langle x_1(t) n_{11}(u) \rangle^0 - \mu k \langle x_1(t) x_1(u) \rangle^0  \right],
\end{equation}
\end{widetext}
where $x_1$ corresponds to the coordinate of the first active particle and $n_{11}$ represents the component of its orientation vector along $X$. In addition, from the steady-state solutions of Eqs. (\ref{eq:ABMmodel})
\begin{widetext}
\begin{equation}\label{eq:solABMmodel}
    x_i(t) = v \int_{-\infty}^t dt' \, e^{- \mu k (t - t')} n_{1i}(t') + \int_{-\infty}^t dt' \, e^{- \mu k (t - t')} \chi_{i}(t'),
\end{equation}
\end{widetext}
and Eqs. (\ref{eq:whitenoise2}) and (\ref{eq:persistorientation}), it can be readily shown that the correlations in Eqs. (\ref{eq:correlkernel1}) and (\ref{eq:correlkernel2}) are explicitly given by
\begin{widetext}
\begin{equation}\label{eq:correlxx}
    \langle x_1(t) x_1(s)\rangle^0 = \frac{k_B T}{k} e^{ - \mu k |t-s| } + \frac{v^2}{3 \left[(\mu k)^2 - D_r^2 \right]} \left(e^{-D_r |t-s|} - \frac{D_r}{\mu k} e^{-\mu k |t-s|}\right),
\end{equation}
\begin{equation}\label{eq:correlxn}
    \langle x_1(t) n_{11}(s)\rangle^0 =\begin{cases}
    \frac{v}{3(\mu k + D_r)}e^{- \mu k (t-s)} + \frac{v}{3(\mu k - D_r)} \left[ e^{-D_r(t-s)} -e^{- \mu k (t-s) } \right]
    , & t \ge s,\\
    \frac{v}{3( \mu k + D_r)}e^{-D_r (s-t)}, & t < s,
  \end{cases}
\end{equation}
\end{widetext}
where it has been assumed that $\langle n_{11}(t) n_{11}(s) \rangle = \langle n_{21}(t) n_{21}(s) \rangle = \langle n_{31}(t) n_{31}(s) \rangle = \frac{1}{3} \langle \mathbf{n}_1(t) \cdot \mathbf{n}_1(s) \rangle $ by isotropy. By substituting the explicit forms of Eqs. (\ref{eq:correlkernel1}) and (\ref{eq:correlkernel2}) using Eqs. (\ref{eq:correlxx}) and (\ref{eq:correlxn}) into Eq. (\ref{eq:kernelact}), we find the following expression for the memory kernel $\Gamma_A(t-s)$
\begin{equation}\label{eq:activekernel}
    \Gamma_A(t-s) = N k e^{- \mu k (t-s)},
\end{equation}
where $t \ge s$. On the other hand, the noise term in Eq. (\ref{eq:GLE2}), $\zeta(t) = - k \sum_{i=1}^N \left[\langle x_i(t) \rangle - x_i(t) \right]$, has zero mean by definition: $\langle \zeta(t) \rangle = 0$. If $N \gg 1$, it becomes Gaussian by virtue of the central limit theorem. Up to quadratic order in $k/\kappa$, its autocorrelation function is
\begin{widetext}
\begin{eqnarray}\label{eq:correlnoise}
    \langle \zeta(t) \zeta(s) \rangle & = & \langle G_X(t) G_X(s) \rangle - \langle G_X(t) \rangle  \langle G_X(s) \rangle \nonumber\\
    & = &  N k^2 \langle \langle x_1(t) x_1(s)  \rangle^0 + \mathcal{O}\left(\frac{k}{\kappa}\right)^3 ,\nonumber\\
    & \approx & N k_B T k e^{- \mu k |t - s|} + \frac{N k v^2}{3\left[ ( \mu k)^2 - D_r^2 \right]} \left( k e^{-D_r |t-s|} - \frac{D_r}{\mu} e^{- \mu k |t-s|} \right).
\end{eqnarray}
\end{widetext}
Hence, in the limit $k \ll N^{-1} \kappa$, Eq. (\ref{eq:GLE2}) can be written as
\begin{eqnarray}\label{eq:GLE_active}
    	\int_{-\infty}^t ds \, \Gamma(t-s) \frac{d}{ds} X(s) & = & - \left(1-N\frac{k}{\kappa} \right)\kappa X(t)  + \xi(t),\nonumber\\
    	&\approx &  - \kappa X(t)  + \xi(t).
\end{eqnarray}
Note that this choice of $k$ guarantees that the stiffness of the total harmonic confinement acting on the PBP is $(1-Nk/\kappa)\kappa\approx \kappa$, in such a way that $\kappa$ is a well-defined control parameter during a quasi-static Stirling cycle. Moreover, in Eq. (\ref{eq:GLE_active}), the complete memory kernel accounting for the total friction exerted on the PBP  due to its interaction with the solvent molecules and the ABPs is
\begin{equation}\label{eq:totalfrictionkernel}
    \Gamma(t-s) = 2\gamma_{\infty} \delta(t-s) + \Gamma_A(t-s),
\end{equation}
with $\Gamma_A(t-s)$ given by (\ref{eq:activekernel}), whereas the total stochastic force acting on the PBP is
\begin{equation}\label{eq:totalstochforce}
    \xi(t) = \zeta(t) + \zeta_{\infty}(t),
\end{equation}
which becomes Gaussian for $N \gg 1$, with zero mean, $\langle \xi(t) \rangle = 0$, and autocorrelation function
\begin{widetext}
\begin{equation}\label{eq:correltotalnoise}
    \langle \xi(t) \xi(s) \rangle = k_B T \Gamma(|t-s|) + \frac{N k v^2}{3\left[ ( \mu k)^2 - D_r^2 \right]} \left( k e^{-D_r |t-s|} - \frac{D_r}{\mu} e^{- \mu k |t-s|} \right).
\end{equation}
\end{widetext}

\end{document}